\documentclass[10pt, prd, twocolumn]{revtex4}

\usepackage{amsmath}
\usepackage{amssymb}
\usepackage{graphicx}
\usepackage{bm} 

\usepackage[dvipsnames]{xcolor}

\usepackage[caption=false]{subfig}
\begin{document}

\title{Thermodynamics of $d$-dimensional Schwarzschild black holes in
the canonical ensemble}



\author{Rui Andr\'{e}}
\email{rui.andre@tecnico.ulisboa.pt}
\affiliation{Centro de Astrof\'{\i}sica e Gravita\c c\~ao  - CENTRA,
Departamento de F\'{\i}sica, Instituto Superior T\'ecnico - IST,
Universidade de Lisboa - UL, Av. Rovisco Pais 1, 1049-001
Lisboa, Portugal}
\newcommand{\e}[0]{\varepsilon}
\newcommand{\dd}[0]{\partial}

\setlength{\textheight}{25cm}

\author{Jos\'{e} P. S. Lemos}
\email{joselemos@tecnico.ulisboa.pt}
\affiliation{Centro de Astrof\'{\i}sica e Gravita\c c\~ao - CENTRA,
Departamento de F\'{\i}sica, Instituto Superior T\'ecnico - IST,
Universidade de Lisboa - UL, Av. Rovisco Pais 1, 1049-001 Lisboa,
Portugal}

\begin{abstract}

We study the thermodynamics of a $d$-dimensional Schwarzschild black
hole, also known as a Schwarzschild-Tangherlini black hole, in the
canonical ensemble. This generalizes York's formalism, which has been
initially applied to four dimensions and later to five dimensions, to
any number $d$ of dimensions. The canonical ensemble, characterized by
a cavity of fixed radius $r$ and fixed temperature $T$ at the
boundary, allows for two possible black hole solutions in thermal
equilibrium, a smaller black hole and a larger black hole. In four and
five dimensions, these solutions have a direct exact form, whereas in
an arbitrary number of dimensions, one is compelled to resort to
approximation schemes or numerical calculations.  From the Euclidean
action and the path integral approach, we obtain the free energy, the
thermodynamic energy, the thermodynamic pressure, and the entropy, of
the black hole plus cavity system. The entropy of the system is given
by the Bekenstein-Hawking area law.  The analysis of the heat capacity
of the system shows that the smaller black hole is in unstable
equilibrium and the larger black hole is in stable equilibrium.  The
$d$-dimensional photon sphere radius divides the stability criterion.
Indeed, 
if the cavity's radius is larger than the photon sphere radius, and so
the black hole is small, the system is unstable, if the cavity's
radius is smaller than the photon sphere radius, and so the black hole
is large, the system is stable.  To study perturbations on the system,
a generalized free energy function is obtained that also allows one to
understand the possible phase transitions between classical hot flat
space and the black holes. The Buchdahl radius, that appears
naturally in the general relativistic study of star structure, also
shows up in our context, the free energy is zero when the cavity's
radius has the $d$-dimensional Buchdahl radius value.  Then, if the
cavity's radius is larger than the Buchdahl radius, classical hot flat
space phase cannot make a phase transition to a black hole phase, and
if smaller, classical hot flat space can nucleate a black hole.
The roles of both the photon sphere and the Buchdahl limit are
present for
every dimension $d$, indicating that, besides their known role in
dynamics, these radii also play a role in the thermodynamics of
gravitational systems.  The close link between
the canonical analysis performed and the direct perturbation of the
path integral
is also pointed out.
Since hot flat space is a quantum system made purely
of gravitons, if only gravitation is considered, it is of great
interest to compare the $d$-dimensional free energies of quantum hot
flat space and the stable black hole to find for which ranges of $r$
and $T$, the quantities that characterize the canonical ensemble, one
phase predominates over the other.  Phase diagrams for a few different
dimensions are displayed. The density of states at a given energy is
found through an inverse Laplace transformation giving back the
entropy of the stable black hole.  Several side calculations and
further deliberations are performed, namely, the calculation for the
approximate expressions for the canonical ensemble black hole horizon
radii, a brief study of the photon orbit radius and the Buchdahl
radius in the $d$-dimensional Schwarzschild solution, a connection to
the thermodynamics of thin shells in $d$ spacetime dimensions which are
systems that are also apt to a rigorous thermodynamic study, a
presentation of quantum hot flat space in $d$ spacetime dimensions as
a thermodynamic system, an analysis of classical hot flat space in $d$
spacetime dimensions as a product of quantum hot flat space with the
black hole transitions and the corresponding phase diagrams for a few
different dimensions, and a synopsis with the relevance of the work.
It is still worth mentioning that the comparison of the thermodynamics
of $d$-dimensional Schwarzschild black holes and classical hot flat
space in the canonical ensemble with the thermodynamics of spherical
thin shells in $d$ dimensions yields a striking direct matching
between the two systems, most notably that the photon sphere radius
appears here as a thermodynamic stability divisor in both systems, and
the Buchdahl radius that appears on thermodynamic grounds for
canonical black holes appears also as a thermodynamic and as a
dynamical radius for thin shells.

\end{abstract}

\keywords{}
\maketitle



\newpage
\centerline{}
\newpage
\centerline{}
\newpage

\section{Introduction}
\label{Intro}

Black holes are physical systems that possess thermodynamic
properties. The path-integral approach to quantum gravity is a
powerful technique that when applied to black holes displays them
clearly as thermodynamic systems.  In this approach the geometry of a
four-dimensional Schwarzschild black hole, say, is Euclideanized and
its temperature is fixed by the correct period in the imaginary time
putting the black hole in a state of equilibrium with a heat bath at
the prescribed temperature, the Hartle-Hawking vacuum state
\cite{hartlehawking}.  The black hole entropy $S$ can then be found to
be $S=\frac14A_+$ in Planck units, where area $A_+$ is the event
horizon area.  This entropy stems from the contribution of the
classical Euclidean Einstein action of the black hole metric to the
partition function and its cause is the nontrivial topology with a
nonzero Euler characteristic of the Euclidean four-dimensional
Schwarzschild black hole, in contrast to ordinary quantum field
theories, where the classical contribution to the action is absorbed
into the normalization of the functional integral \cite{hawking}.  It
is of great interest to extend this approach to $d$-dimensional
Schwarzschild black holes.
The generalization of the Hartle-Hawking vacuum setting to
$d$-dimensional Schwarzschild black holes has been done in  \cite{mhewitt}.
Moreover, the understanding that the black hole entropy in $d$
dimensions, with $d\geq4$, comes from topological considerations,
specifically, the Euler characteristic of the two- dimensional plane
spanned by the Euclidean time and radial spatial coordinate, was
performed in \cite{btz1994}.

With the path-integral approach in hand, York understood that the
correct setting to study black thermodynamics, in particular a
four-dimensional Schwarzschild black hole, was to work with the
canonical ensemble of statistical mechanics \cite{york1} which
provides a complete description of the thermodynamics of those
systems.  In the canonical ensemble, the black hole is placed inside a
cavity whose boundary has radius $r$ and is at temperature $T$, i.e.,
the cavity is in thermal equilibrium with a heat bath at temperature
$T$.  The Euclidean action for the system shows that the instanton
solution admits two boundary configurations, i.e., there are two black
hole solutions for the canonical boundary data.  One solution yields a
small black hole inside a large cavity in thermal, but unstable,
equilibrium, which was the system studied in great detail in
\cite{gpy} that had been previously studied in
\cite{hartlehawking,hawking}. The other solution yields a large black
hole inside a cavity with a size of the same order of the black hole,
in thermal and stable equilibrium, which was studied in \cite{allen}.
By using the canonical ensemble, and showing there are stable
configurations, the thermodynamics of black holes is then unified with
a proper setting.
The canonical ensemble path-integral approach can be extended to more
complex systems, as has been done for electrically charged black holes
in the grand canonical ensemble \cite{york2}, for black holes in
anti-de Sitter spacetimes \cite{pecalemos}, and even for matter
configurations \cite{zaslavskii1} or matter plus black hole systems
\cite{yorkmartinez}.
In higher dimensions, York's formalism can also be developed.  The
five-dimensional Schwarzschild black hole has shown to be of
particular interest, because the exact solutions for the instantons
take a simple form which allows for an approach with fewer
approximations than those used originally, with the smaller unstable
solution and the larger stable solution being found exactly
\cite{andrelemosd5}. Moreover, the stable and unstable thermodynamic
modes of a $d$-dimensional Schwarzschild black hole have been studied
in detail in \cite{GregRoss}, see also \cite{reallbranes,lu}.
In this work, we generalize the four- and five-dimensional canonical
ensemble path integral approach for a spherical symmetric black hole
in $d$ spacetime dimensions, enabling us to extract
intrinsic features that might arise.

Now, another gravitational system that can be handled in pure
thermodynamic grounds is a spherical thin shell that separates a
Minkowski interior from some exterior spacetime. Fixing the
temperature on the shell, and given a well-prescribed first law of
thermodynamics at the shell, a powerful thermodynamic formalism can be
developed that gives the entropy and the stability of the shell. In
four dimensions, for a shell with a Schwarzschild exterior,
the problem
was treated in \cite{Martinez}, and for a shell with a
Reissner-Nordstr\"om exterior, the problem was treated in
\cite{lemosetal}.  The study of thermodynamics of thin shells in $d$
dimensions with a Schwarzschild exterior was solved in \cite{dshells}.
We are thus led to compare here the $d$-dimensional black hole
in the canonical ensemble studied in this work
with the $d$-dimensional thin matter
shells in the thermodynamic setting
studied in \cite{dshells}.

There are some results that will be used.  In four dimensions, the
solution we are interested in is the Schwarzschild solution.  An
analysis on the quasilocal energy of spherical spacetimes that
has applications
in thermodynamic problems was done in \cite{by}. The photonic radius,
the radius where the photons have circular orbits, in Schwarzschild in
four dimensions is given by $r=\frac32 r_+$ where $r_+$ is the
gravitational radius, and since $r_+=2m$, one also can write $r=3m$,
where $m$ is the spacetime mass. This special radius also appears in
the thermodynamic study of the black hole in the canonical ensemble,
as York noticed. The Buchdahl radius, i.e., the radius for the maximum
compactness of a general relativistic star \cite{buchdahl},
or of a general relativistic thin shell
under certain conditions
\cite{andreasson}, is given by 
$r=\frac98 r_+$ where $r_+$ is the
gravitational radius, and since $r_+=2m$, one also can write
$r=\frac94m$, where again
$m$ is the spacetime mass. This special radius
also appears in the thermodynamic study of the black hole in the
canonical ensemble, as we noticed here.  When studying the black hole
in the canonical ensemble, one also needs the thermodynamic properties
of a radiation gas in four spacetime dimensions as given in any book
in thermodynamics.
In higher $d$ dimensions, the solution we are interested in is the
$d$-dimensional Schwarzschild solution \cite{tangherlini}, also called
Schwarzschild-Tangherlini or simply Tangherlini solution.  Quasilocal
energy on higher-dimensional spacetimes has not been performed but
certainly the results are maintained.  The photonic radius in
Schwarzschild in $d$ dimensions is given in \cite{monteiro}.  This
special radius also appears in the thermodynamic study of the
$d$-dimensional black hole in the canonical ensemble.  The Buchdahl
radius for most compactness of a $d$-dimensional star is given in
\cite{wright} and for a shell in a $d$-dimensional spacetime we give
here.  This special radius also appears in the thermodynamic study of
the black hole in the canonical ensemble, as we noticed here. 
When
studying the black hole in the canonical ensemble, one needs to use the
thermodynamic properties of a radiation gas in $d$ spacetime
dimensions as given in \cite{landsberg}.

The paper is organized as follows.  
In Sec.~\ref{secCavity}, we
prepare the cavity at a fixed radius $r$ and temperature $T$ at
the cavity's wall.
Inside the cavity, for the Schwarzschild-Tangherlini metric,
we look for the black hole solutions which satisfy thermal equilibrium
with the cavity's wall.
The section is split in two parts, where we find an expression for the
smaller black hole first, followed by the larger one.
We also derive the Euclidean Einstein-Hilbert action
for a $d$-dimensional Schwarzschild black hole as a function of the
cavity's radius and temperature.
In Sec.~\ref{secThermo}, from the action, we
derive all the thermodynamic quantities associated to the black hole
plus cavity system, particularly, the thermodynamic
energy, pressure, and
entropy, along with the first law of thermodynamics for the system.
In Sec.~\ref{secStability}, we find the heat capacity for the system,
which is crucial in identifying the thermal stability of the solutions.
In Sec.~\ref{secFreeEnergyFunc}, with the free energy function being
the thermodynamic potential of the canonical ensemble proportional
to the action, we can better interpret possible state transitions
inside the cavity, discussing the possibility of black hole nucleation,
or even the transition from a black hole state to flat space.
In Sec.~\ref{sa}, we address and comment on the relationship between
the action functional to second order, and thermodynamics and thermal
stability.
In Sec.~\ref{secGround}, we directly compare the free energy of
$d$-dimensional quantum hot flat space with the free energy of the
stable black hole. With this, we can identify the conditions for each
of these states being the ground state of the canonical ensemble,
i.e., with the lowest free energy, or when the ground state is a
superposition of both, when they have the same free energy.  In
Sec.~\ref{secDensity}, we compute the density of states from the
partition function for the stable black hole solution, which in turn
leads to an alternative way of reproducing the area law for the
entropy.
In Appendix~\ref{calculations},
we develop some side calculations.
In Appendix~\ref{phbuch},
we dwell on two important radii that appear
in the canonical ensemble context, the photon orbit radius
and the Buchdahl radius.
In Appendix~\ref{secshells}, we establish the relationship between 
the thermodynamics of black holes in a cavity
in $d$ dimensions and
the thermodynamics of  thin matter
shells in $d$ dimensions.
In Appendix~\ref{secApB}, we derive the generalized $d$-dimensional
free energy and action
for quantum hot flat space, along with the thermodynamic  
quantities used.
In Appendix~\ref{appchfs},
we study classical hot flat space in $d$ spacetime dimensions as 
the product of quantum hot flat space and
analyze the corresponding black hole
phase transitions for classical hot flat space.
In Appendix~\ref{conc}, we present a synopsis
and further additions.

\vfill

\section{Canonical ensemble for a cavity with a black hole inside:
Temperature, the Euclidean Einstein action, and the action functional
or partition function for a $d$-dimensional Schwarzschild black hole}
\label{secCavity}

\subsection{Cavity in $d$ dimensions and
the canonical temperature}

\subsubsection{Generics and temperature of the canonical ensemble}

In the canonical ensemble
of a spherical symmetric thermodynamic
system,
we fix the
radius $r$ of the cavity's boundary and the
local temperature $T$ at the cavity's boundary. We also
define the 
inverse temperature $\beta= \frac1T$, 
which is a useful parameter, so that the independent variables we will
work with can either be $T$ and $r$ or $\beta$ and $r$.  Throughout
the paper, we set the speed of light $c$, the gravitational constant
$G$, the Planck constant $\hbar$, and the Boltzmann constant $k_B$ to
unity, i.e., $c=1$, $G=1$, $\hbar=1$, and $k_B=1$.  As a consequence,
the Planck length is given by $l_P=1$, and the Planck temperature
is given by
$T_P=1$.

The black hole solution inside the cavity follows from the
$d$-dimensional Schwarzschild solution, also called the
Schwarzschild-Tangherlini solution, with line element given by
\begin{equation}
\label{metric}
ds^2 = \left( 1- \frac{{r_+}^{d-3}}{r^{d-3}} \right) dt^2
+
\dfrac{dr^2}{1- \frac{{r_+}^{d-3}}{r^{{d-3}}}}
+ r^2 d\Omega_{d-2}^2,
\end{equation}
where $t$ is Euclidean time, $r$ is the coordinate radius, and
$d\Omega_{d-2}^2 = d\theta_1^2+ \sum_{k=2}^{d-2} \left(
\prod_{j=1}^{k-1}
\sin^2 \theta_j \right) d\theta_k^2$ is the line element on the
$(d-2)$-sphere, with the $\theta_k$ being its angles.
We are using the symbol $r$ with two different meanings. One $r$ is
the coordinate radius $r$ of Eq.~(\ref{metric}).
The other $r$ is the cavity's
radius $r$.  The coordinate radius $r$ will disappear soon and will
not be mentioned anymore, so there is no possibility of confusion.
In $d$
dimensions. the gravitational radius, being also the event horizon
radius when there is a black hole, $r_+$, and the spacetime mass $m$,
sometimes called the ADM mass, are related by
$r_+^{d-3}=\frac{16\pi}{(d-2)\Omega_{d-2}}m$,  where $\Omega_{d-2}
= \frac{2\pi^{\frac{d-1}{2}}}{\Gamma\left(\frac{d-1}{2}\right)}$ is
the solid angle in a spherical $d$-dimensional spacetime.  Clearly,
from Eq.~(\ref{metric}), we have to impose $d\geq4$, so that the
canonical ensemble here is valid for a four- or higher-dimensional
spacetime.

The Euclidean metric in Eq.~(\ref{metric}) describes the spacetime of
an Euclidean black hole outside the horizon, i.e., the coordinate $r$
obeys $r\geq r_+$, provided that the conical singularity at $r=r_+$ is
removed by setting the correct time period to $t$. By redefining the
coordinate $r$ as $r=r_+ + \varepsilon$, with $\varepsilon$ a radial
variable such that $\varepsilon\ll r_+$, and introducing then a new
radial coordinate $\rho = \sqrt{\frac{4r_+ \varepsilon}{d-3}}$, the
metric given in Eq.~(\ref{metric}) reduces to $ds^2= d\rho^2 + \rho^2
\left(\frac{2r_+}{d-3}\right)^2 dt^2$.  So, in order to have no
conical singularities, $t$ must have a period, which we will denote by
$\beta_\infty$, given by $\beta_{\infty}=\frac{4\pi r_+}{d-3}$.  This
$\beta_\infty$ is the inverse Hawking temperature. So, the
Hawking temperature $T_H$, i.e., the temperature at infinity
for $d$-dimensional black holes,
is
$T_H=\frac1{\beta_{\infty}}=\frac{d-3}{4\pi r_+}$.
Now, the Tolman temperature says that the temperature at some position
$r$ is the temperature at infinity blueshifted to $r$. From now on $r$
denotes always the radius of the cavity.
So, in order that there is thermal equilibrium
between the black hole and the cavity at $r$, the temperature, or its
inverse $\beta$, at $r$, must satisfy the Tolman formula. Therefore,
$\beta = \beta_\infty
\sqrt{1-\frac{r_+^{d-3}}{r^{d-3}}}$, or using $
\beta_{\infty}=\frac{4\pi r_+}{d-3}$ we obtain
\begin{equation}\label{tolman}
\beta = \frac{4\pi r_+}{d-3} \sqrt{1-\frac{r_+^{d-3}}{r^{d-3}}}.
\end{equation}
Since
\begin{equation}
\label{beta}
T=\frac1\beta,
\end{equation}
in terms of $T$,
Eq.~(\ref{tolman}) is
\begin{equation}\label{polynomial}
\left(\frac{r_+}{r} \right)^{d-1}-\left(\frac{r_+}{r} \right)^2+
\left( \frac{d-3}{4\pi r T} \right)^2 = 0\,.
\end{equation}
Equation~(\ref{polynomial}) is a polynomial equation with its order set by
$d$.  Exact solutions for $d=4$ and $d=5$ were obtained
in~\cite{york1} and~\cite{andrelemosd5}, respectively.
In general,
for $d\geq6$, one is compelled to resort to approximation
schemes or numerical calculations to
solve Eq.~(\ref{polynomial}), although in some
dimensions, an exact, although contrived, analysis
might be performed,
noting 
that for odd $d$ Eq.~(\ref{polynomial}) can have its order reduced by
solving for $\left(\frac{r_+}{r}\right)^2$.

To deal with
Eq.~(\ref{polynomial}), we note that the cavity radius $r$ has
range $r_+\leq r<\infty$, i.e., $0\leq
\frac{r_+}{r} \leq 1$.  Let us also write the left-hand side of
Eq.~(\ref{polynomial}) as a function $f\left(\frac{r_+}{r}\right)$
such that $f\left(\frac{r_+}{r}\right)=\left(\frac{r_+}{r}
\right)^{d-1}- \left(\frac{r_+}{r} \right)^2+ \left(\frac{d-3}{4\pi r
T}\right)^2$.  Then, at $\frac{r_+}{r}=0$, one has
$f(0)=\left(\frac{d-3}{4\pi r T}\right)^2$, and at $\frac{r_+}{r}=1$,
one has $f(1)=\left(\frac{d-3}{4\pi r T}\right)^2$, so the extreme
points of the interval have the same positive value.
From the first
derivative of $f\left(\frac{r_+}{r}\right)$, one finds that it has a
unique extremum, in fact a minimum, at $\frac{r_+}{r}=\left(
\frac{2}{d-1}\right)^{\frac1{d-3}}$, so that $f_{\rm
min}=f\left(\left( \frac{2}{d-1}\right)^{\frac1{d-3}}\right)$.
Moreover, the second derivative at this minimum of
$f\left(\frac{r_+}{r}\right)$ is always positive.  So,
$f\left(\frac{r_+}{r}\right)$ starts at $\frac{r_+}{r}=0$ with value
$f(0)=\left(\frac{d-3}{4\pi r T}\right)^2$ positive, decreases up to
$\frac{r_+}{r}=\left( \frac{2}{d-1}\right)^{\frac1{d-3}}$, where it
has a minimum value $f_{\rm min}=f\left(\left(
\frac{2}{d-1}\right)^{\frac1{d-3}}\right)$, and increases back up to
$\frac{r_+}{r}=1$ with value $f(0)=\left(\frac{d-3}{4\pi r
T}\right)^2$ positive.  Clearly, there are solutions to
Eq.~(\ref{polynomial}) only if $f_{\rm min}= f\left(\left(
\frac{2}{d-1}\right)^{\frac1{d-3}}\right)\leq0$.  Since
Eq.~(\ref{polynomial}) has only one minimum, there will be in general
two solutions that degenerate into one only when the equality in the
latter equation holds.

In brief, Eq.~(\ref{polynomial}) only has solutions if $f_{\rm
min}\left( \frac{r_+}{r} \right)\leq 0$, i.e., $f\left(\left(
\frac{2}{d-1}\right)^{\frac1{d-3}}\right)\leq 0$.  So, the condition
for the canonical ensemble at fixed $r$ and $T$ to have black hole
solutions $\frac{r_+}{r}$ is from Eq.~(\ref{polynomial})
\begin{equation}\label{bhcondition}
\pi r T \geq \frac{d-3}{4}\left[
\left( \frac{2}{d-1} \right)^{\frac{2}{d-3}}
-\left( \frac{2}{d-1} \right)^{\frac{d-1}{d-3}}\right]^{-1/2}\,.
\end{equation}
There will indeed be two possible black hole
solutions,
$\frac{r_{+1}}{r}$ and 
$\frac{r_{+2}}{r}$, and when 
the equality holds there is only one black hole solution,
$\frac{r_{+1}}{r}=\frac{r_{+2}}{r}$.

Let us see some further properties of Eq.~(\ref{bhcondition}). 
Equation~(\ref{bhcondition}) gives that the minimum value 
that $\pi r T$ can
take is given by the number of dimensions only, a property that can
be clearly seen when one treats the $d$-dimensional case
generically.  Equation~(\ref{bhcondition}) also shows that as $d$
increases, the minimum value of $\pi r T$ also increases.  Indeed, for
$d=4$, the threshold value for the existence of a black hole is $\pi r
T= \frac{3\sqrt3}{8}$, or $\pi r
T=0.650$ approximately.  For
$d=5$, the threshold value for the existence of a black holes is $\pi r
T = 1$. For $d\geq 6$, Eq.~(\ref{bhcondition}) gives that the
threshold value is always larger than $1$.
Given Eq.~(\ref{bhcondition}), we need from  Eq.~(\ref{polynomial})
to find an expression 
for the two black hole solutions, i.e., for 
$\frac{r_{+1}}{r}$ and 
$\frac{r_{+2}}{r}$.
Clearly, for $\pi r T \gg 1$,
Eq.~(\ref{polynomial}) reduces to
$(\frac{r_+}{r})^{d-1}-(\frac{r_+}{r})^2=0$, so in this case, 
the two black hole 
solutions will be expansions around $\frac{r_+}{r}=0$ and
$\frac{r_+}{r}=1$.
We now turn to find approximate  solutions for 
$r_{+1}$ and $r_{+2}$.

\subsubsection{Smaller black hole solution $r_{+1}$
and the larger black hole solution $r_{+2}$}
\label{Small}

\noindent
{\it Smaller black hole solution $r_{+1}$:}
\vskip 0.1cm

\noindent
To find the smaller black hole solution $\frac{r_{+1}}r$ around
$\frac{r_+}r=0$, we make a Taylor expansion
and write
$\frac{r_{+1}}r =\frac{r_{+1}}r(\pi r T)$ as
$\frac{r_{+1}}r\left(\pi r T\right) = \frac{a_1}{\pi r T}
+\frac{a_2}{ \left( \pi r T \right)^2} 
+ ...$,
where the $a_i$ are constants to be determined.
Equating carefully
power by power
this expansion in 
Eq.~(\ref{polynomial}), one finds,
see Appendix~\ref{calculations}, 
\begin{equation}\label{bh1}
r_{+1} = r\, \left( \frac{d-3}{4 \pi r T} +
\frac12 \left(\frac{d-3}{4 \pi r T}\right)^{d-2} + 
\mathcal O\left(\frac1{(\pi r T)^{d-1}}\right) \right)\,.
\end{equation}
This is the smaller black hole solution $\frac{r_{+1}}{r}$ for large
$T$.

\vskip 0.3cm
\noindent
{\it Larger black hole solution $r_{+2}$}
\vskip 0.1cm

\noindent
To find the larger black hole solution $\frac{r_{+2}}r$
around $\frac{r_+}r=1$,
we make a Taylor expansion
and write
$\frac{r_{+2}}r =\frac{r_{+2}}r(\pi r T)$ as
$\frac{r_{+2}}r (\pi r T) = 1 + 
\frac{b_1}{\pi r T} + \frac{b_2}{(\pi r T)^2}+ ... $,
where the $b_i$ are constants to be determined.
Equating carefully
power by power
this expansion in 
Eq.~(\ref{polynomial}) one finds,
see Appendix~\ref{calculations}, 
\begin{equation}\label{bh2}
r_{+2} = r\left(1 - \frac{d-3}{16\left( \pi r T\right)^2} 
+ \mathcal O\left( \frac1{(\pi r T)^4} \right) \right).
\end{equation}
This is the larger black hole solution $\frac{r_{+2}}{r}$ for large
$T$.

\vskip 0.3cm
\noindent
{\it Equal radius black hole solution $r_{+1}=r_{+2}$}
\vskip 0.1cm

\noindent
Now, there is a $T$, not large where the two black holes have equal
horizon radii. This happens when the equality in
Eq.~(\ref{bhcondition}) holds,
i.e., $\pi r T = \frac{d-3}{4}\left[
\left( \frac{2}{d-1} \right)^{\frac{2}{d-3}}
-\left( \frac{2}{d-1} \right)^{\frac{d-1}{d-3}}\right]^{-1/2}$.
In this case,
there is only one black hole solution for
Eq.~(\ref{polynomial}), namely,
\begin{equation}\label{r+1=r+2}
\frac{r_{+1}}{r}=\frac{r_{+2}}{r} =
\left( \frac{2}{d-1} \right)^\frac1{d-3}\,.
\end{equation}
This means that the cavity's radius $r$ is located at the black hole's
photon sphere, since the 
photon sphere radius is given by $r_{\rm ph}=
\left( \frac{d-1}{2} \right)^\frac1{d-3}r_+$,
see \cite{monteiro} for the black hole photon sphere in
$d$ dimensions, see also Appendix~\ref{phbuch}.

\vskip 0.3cm
\noindent
{\it The Full solution for $r_{+1}$ and $r_{+2}$:}
\vskip 0.1cm

\begin{figure}[h]
\centering
\includegraphics[width=0.48\textwidth]{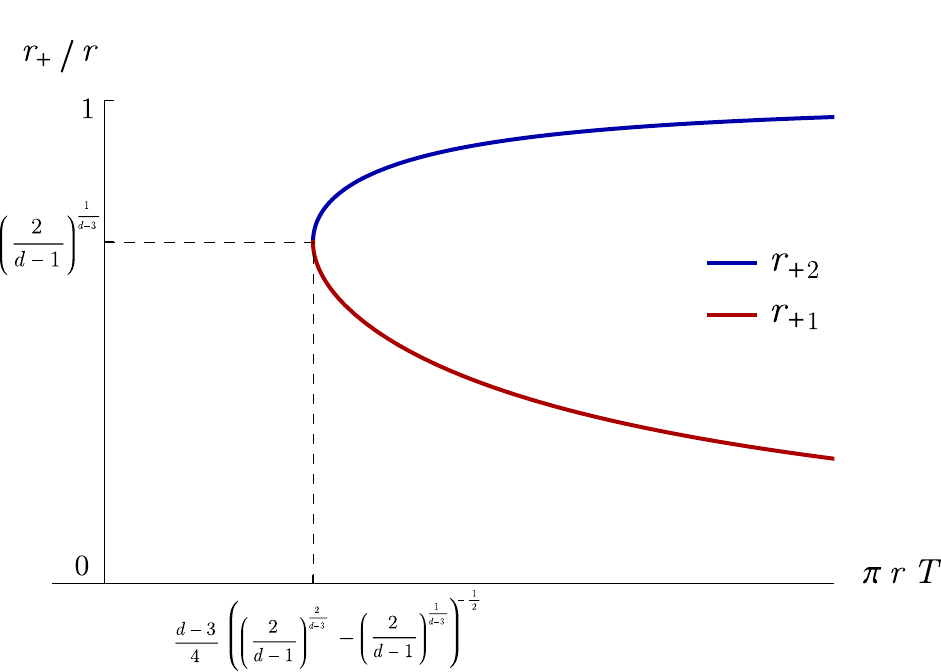}
\caption{
The two black hole solutions $r_{+1}$ and $r_{+2}$ are shown in a plot
$\frac{r_{+1}}{r}$ and $\frac{r_{+2}}{r}$, with $r$ being the cavity
radius, as a function of $\pi r T$, with $\pi r T$ obeying $\pi r T
\geq \frac{d-3}{4}\left[ \left( \frac{2}{d-1} \right)^{\frac{2}{d-3}}
-\left( \frac{2}{d-1} \right)^{\frac{d-1}{d-3}}\right]^{-1/2}$.  The
two solutions coincide when $\frac{r_{+1}}{r}=\frac{r_{+2}}{r} =
\left( \frac{2}{d-1} \right)^\frac1{d-3}$, represented by a point at
the intersection of the dashed lines.  The corresponding radius $r$ is
the radius of the photon sphere of a $d$-dimensional black hole,
$r_{\rm ph}= \left( \frac{d-1}{2} \right)^\frac1{d-3}r_+$.
}
\label{BHs}
\end{figure}

\noindent
In Fig.~\ref{BHs}, the full solution of Eq.~(\ref{polynomial}) is
drawn displaying $r_{+1}$ and $r_{+2}$ as a function of $\pi r T$.
The details are dependent on the dimension $d$ of the spacetime, but
the main features are as shown.

\subsubsection{Location and the area of the cavity}

Another important characterization of the canonical ensemble,
besides its temperature 
is its location
given by the radius $r$ of the cavity's boundary.
In some instances, it is preferable to work with the
cavity's boundary area $A$, which can be given in terms
of $r$ as
\begin{equation}
\label{areacav}
A=\Omega_{d-2} \, r^{d-2}\,,
\end{equation}
with $\Omega_{d-2}$ being the solid angle in a
spherical $d$-dimensional spacetime.

\subsection{Euclidean-Einstein action
and the action functional or the partition function
for a $d$-dimensional Schwarzschild black hole}
\label{secAction}

In the path integral approach to quantum gravity, i.e., the
Hartle-Hawking approach, integration of the Euclidean Einstein action
over the space of metrics $\rm g$ yields the canonical partition
function $Z=\int \mathcal{D}[{\rm g}]\exp(-I [{\rm g}])$ Taking a
black hole solution as the background metric, the leading term in the
expansion will be that of the classical action, specifically,
\begin{equation}\label{partition}
Z={\rm e}^{-I}\,.
\end{equation}
Here, $I$ is the Euclidean Einstein action of the gravitational
system, being the black hole action 
if the system contains a black hole. 

For a $d$-dimensional spacetime
the Euclidean Einstein action $I$ is
\begin{equation}\label{initaction}
I = - \frac{1}{16\pi} \int_\mathcal{M} d^{d}x\sqrt{|g|}R -
\frac1{8\pi} \int_{\partial \mathcal M} d^{d-1}x\sqrt{|h|}[K],
\end{equation}
where $\mathcal M$ is a compact region
of the spacetime and
$\partial \mathcal M$ is its boundary,
$|g|$ is 
the determinant of the  $d$-dimensional spacetime metric
$g_{ab}$, $R$ is the corresponding Ricci scalar,
$|h|$ is the determinant of the
$(d-1)$-dimensional induced  metric on the boundary,
and
$[K]$ is the difference of 
the extrinsic curvature $K$ on the boundary
to
the extrinsic curvature of an equivalent
boundary embedded in
flat space, $K_{\rm flat}$. This subtraction is needed
in order to normalize the  action and the
energy of the 
ensemble.

Given that we are interested
in a vacuum solution, $R=0$, the action of 
Eq.~(\ref{initaction}) reduces to the boundary term.
For the metric
Eq.~(\ref{metric}),
the line element on the boundary $\partial \mathcal M$ for
$r=\rm constant$ is
$ds^2|_{\partial \mathcal M} = 
\left( 1 - \frac{r_+^{d-3}}{r^{d-3}} \right) dt^2
+ r^2 d\Omega_{d-2}^2$. Writing 
$ds^2|_{\partial \mathcal M} = h_{\alpha\beta}dx^\alpha dx^\beta$,
with $\alpha,\beta$
being indices for the time $t$ and the angles $\theta_i$,
one finds that
the
determinant of the induced metric is
$h = \left( 1 - \frac{r_+^{d-3}}{r^{d-3}} \right) r^{2(d-2)}
\prod_{i=1}^{d-2} \sin^{2(d-2-i)}\theta_i$.
The extrinsic curvature of the boundary at $r$ in $d$ dimensions
can be calculated to be 
$K = \frac{d-2}r \sqrt{1- \frac{r_+^{d-3}}{r^{d-3}}} + 
\frac{d-3}{2r\sqrt{1- \frac{r_+^{d-3}}{r^{d-3}}}} 
\left( \frac{r_+}r \right)^{d-3}$.
The flat counterpart can be obtained by setting $r_+=0$,
$K_{\rm flat} = \frac{d-2}{r}$.
To perform the integral in Eq.~(\ref{initaction}),
note that the
coordinates on the boundary, $t$ and $\theta_i$, can be
separated into an integral over the time component, and
an integral over the angles, so that 
$ I= \frac{ [K]}{8\pi}\sqrt{1- \frac{r_+^{d-3}}{r^{d-3}}} 
\int_0^{\beta_\infty}dt \int_{\Omega_{d-2}} r^{d-2}d\Omega_{d-2}$,
where the Euclidean time is integrated over the period 
$\beta_\infty$ defined above, i.e., $
\beta_{\infty}=\frac{4\pi r_+}{d-3}$.
Using $[K]=K-K_{\rm flat}$,
the black hole
Euclidean action as a function of the cavity's boundary 
radius $r$ and the gravitational radius
$r_+$ is then
\begin{align}
I &= \frac{(d-1)\Omega_{d-2}}{4(d-3)}r_+^{d-2} - 
\frac{(d-2)\Omega_{d-2}}{2(d-3)}r_+ r^{d-3} +  \nonumber \\
& + \frac{(d-2)\Omega_{d-2}}{2(d-3)}r_+ r^{d-3}
\sqrt{1- \frac{r_+^{d-3}}{r^{d-3}}}.
\label{actionr+}
\end{align}
In this form, one has that $I=I(r,r_+)$.

Since the thermodynamic variables that fix the
canonical ensemble
are $r$ and $\beta$, or 
equivalently,
$r$ and $T$ if one prefers,
we want to write the action (\ref{actionr+}) 
as a function of $r$ and $\beta$ only, 
$I=I(r,\beta)$.
Noting that 
$r_+=r_+(r,\beta)$, see Eqs.~(\ref{bh1})~and~(\ref{bh2})
and Fig.~\ref{BHs}, one has that Eq.~(\ref{actionr+}) 
can be formally rewritten as
\begin{align}
I(r,\beta) = &\frac{(d-1)
\Omega_{d-2}}{4(d-3)}\left[r_+(r,\beta)\right]^{d-2} - \nonumber \\
&\frac{(d-2)\Omega_{d-2}}{2(d-3)}r_+(r,\beta) r^{d-3}  
 + \frac{(d-2)\Omega_{d-2}}{8\pi T} r^{d-3}\,,
\label{actionr+beta}
\end{align}
with the help of Eqs.~(\ref{tolman}) and~(\ref{beta}) 
for the last term, 
where $r_+$ stands for
$r_{+1}$ and $r_{+2}$,
With the approximation found in Eq.~(\ref{bh1}) for 
$r_{+1}(r,\beta)$, the action for the small black hole is
\begin{equation}\label{actionbh1}
I(r,r_{+1}(r,\beta)) = \frac{\Omega_{d-2} r^{d-2}}{4(d-3)}\left( 
\frac{d-3}{4\pi r T}\right)^{d-2} +  \mathcal O \left( 
\frac{1}{\left( \pi r T \right)^{d-1}}\right),  
\end{equation}
which is always positive.
With the approximation found in Eq.~(\ref{bh2}) for 
$r_{+2}(r,\beta)$, the action for the large black hole is
\begin{align}\label{actionbh2}
I(r,r_{+2}(r,\beta)) &= - \frac{\Omega_{d-2} r^{d-2}}{4}\left( 
1 - \frac{d-2}{2 \pi r T} + 
\frac{(d-2)(d-3)}{16\left(\pi r T\right)^2} \right) +
\\ \nonumber
&+\mathcal O \left( 
\frac{1}{\left( \pi r T \right)^{4}}\right),  
\end{align}
which will be positive for small values of $\pi r T$,
provided they still satisfy the condition for existence of
equilibrium given in Eq.~(\ref{bhcondition}), and will be 
negative for all the other values of $\pi r T$.
From Eq.~(\ref{actionr+}), one can also take that the action
of the larger black hole is
positive for 
$\frac{r}{r_+} > \left( \frac{ (d-1)^2 }{ 4(d-2) } 
\right)^{\frac{1}{d-3}}$
and is
negative for 
$\frac{r}{r_+} < \left( \frac{ (d-1)^2 }{ 4(d-2) } 
\right)^{\frac{1}{d-3}}$.
Since to have a system at all one must
impose $r>r_+$, 
the action exists and is negative for
$1<\frac{r}{r_+} <  \left( \frac{ (d-1)^2 }{ 4(d-2) } 
\right)^{\frac{1}{d-3}}$,
which can only be achieved by the larger black hole
$r_{+2}$.
Thus, in brief, 
the action given in Eq.~(\ref{actionr+}) is zero or positive
for
\begin{equation}\label{buch}
\frac{r}{r_+} \geq \left( \frac{ (d-1)^2 }{ 4(d-2) } 
\right)^{\frac{1}{d-3}}.
\end{equation}
Note that $\left( \frac{ (d-1)^2 }{ 4(d-2) } \right)^{\frac{1}{d-3}}$
sets an important cavity radius $r$ in terms of $r_+$, the
Buchdahl radius, 
as we will discuss below, see also Appendix~\ref{phbuch}.

\section{Thermodynamics}
\label{secThermo}

The statistical mechanics canonical ensemble
setting of black holes is given through
the partition function $Z$ and its action 
$I$ in Eq.~(\ref{partition}), 
where $I$
takes the form
of Eq.~(\ref{actionr+}), or 
Eq.~(\ref{actionr+beta}),
and the connection
to thermodynamics
is made by the relation between $I$ and
the free energy $F$, the relevant 
thermodynamic potential usually used
in the canonical  context.
The needed relation is 
\begin{equation}
\label{FandI}
I=\beta F\,.
\end{equation}
In thermodynamics, the thermodynamic energy $E$ and the entropy $S$ are
also important thermodynamic potentials and the relation between $F$,
$E$, and $S$ is
\begin{equation}
\label{FandETS}
F=E-TS.
\end{equation}

Now, to establish the first law of thermodynamics, we envisage $E$ as
the main thermodynamic potential and assume it to be a function of
the entropy $S$ and the cavity area $A$, $E=E(S,A)$. The first law of
thermodynamics can then be written as
\begin{equation}
\label{firstlaw}
dE = TdS - pdA,
\end{equation}
where $T$ is the thermodynamic variable
conjugated to $S$, i.e., the temperature,
that has to be found as an 
equation of state of the form $T=T(S,A)$,
and $p$ is the thermodynamic variable
conjugated to $A$, i.e., the tangential pressure
or
the pressure perpendicular to the
cavity radius $r$,
that has to be found as an 
equation of state of the form $p=p(S,A)$.
All quantities, $E$, $T$, $S$, $p$, and $A$, are local
or quasilocal quantities defined at the cavity's location.  To perform
calculations directly with the action $I$ given in
Eq.~(\ref{actionr+}), or Eq.~(\ref{actionr+beta}), one changes
variables in in the first law Eq.~(\ref{firstlaw}) to the variable $F$
and then to $I$ using Eq.~(\ref{FandETS}) followed by
Eq.~(\ref{FandI}).  We have $dF=dE-TdS-SdT$ and $dI= \beta dF +
Fd\beta$, so that the first law can be written as
\begin{equation}
\label{firstlawnew}
dI=Ed\beta-p \beta dA\,,
\end{equation}
i.e., $I=I(\beta,A)$.
Then,
$E$, $p$ and $S$ are given by
\begin{align}
E &= \left( \frac{\partial I}{\partial \beta} \right)_A, 
\label{E} \\
p &= - \frac{1}{\beta} \left( \frac{\partial I}{\partial A}
\right)_\beta, 
\label{p} \\
S &= \beta E - I\,,
\label{S}
\end{align}
respectively.
We can now find
$E$, $p$, and $S$.

To obtain the thermodynamic $E$, we have to perform
the derivative
$\left( \frac{\partial I}{\partial \beta} \right)_A$.
It is simpler to use the cavity radius $r$
instead of its area $A$, which can be done
through Eq.~(\ref{areacav}).
If $I$ is seen as $I=I(r,\beta)$, then
$dI=\left( \frac{\partial I}{\partial \beta} \right)_r
d\beta+
\left( \frac{\partial I}{\partial r} \right)_\beta dr
$.
If $I$ is seen as $I=I(r,r_+)$ then
$dI=\left( \frac{\partial I}{\partial r_+} \right)_r
dr_++
\left( \frac{\partial I}{\partial r} \right)_{r_+} dr
$. Equating these two equations at constant $r$
one obtains
$\left( \frac{\partial I}{\partial \beta} \right)_r = 
\frac{(\partial I / \partial r_+)_r}{(\partial \beta /
\partial r_+)_r}$.
Using Eqs.~(\ref{tolman}) and~(\ref{actionr+})
in Eq.~(\ref{E}) 
yields
\begin{equation}\label{Efinal}
E =  \frac{(d-2)\Omega_{d-2}\,r^{d-3}}{8\pi} 
\left(1- \sqrt{1- \frac{r_+^{d-3}}{r^{d-3}}} \right).
\end{equation}
The total thermodynamic energy is larger than the
spacetime mass $m$, and one can decompose the spacetime mass as the
thermodynamic energy inside the cavity minus its gravitational 
binding energy, i.e. 
$m = E - \frac{4\pi E^2}{(d-2)\Omega_{d-2}\, r^{d-3}}$, where 
$r_+^{d-3}=\frac{16\pi}{(d-2)\Omega_{d-2}}m$ has been used.
This thermodynamic energy $E$ is also a quasilocal energy \cite{by}.

To obtain
the thermodynamic pressure $p$
note that 
$\left(\frac{\partial I}{\partial r} \right)_\beta =
\left(\frac{\partial I}{\partial r} \right)_{r_+} - 
\left(\frac{\partial I}{\partial \beta} \right)_r
\left(\frac{\partial \beta}{\partial r} \right)_{r_+}$,
where again it is simpler to use the cavity radius $r$
instead of its area $A$, which can be done
through Eq.~(\ref{areacav}).
Using Eqs.~(\ref{tolman}) and~(\ref{actionr+})
in Eq.~(\ref{p})
yields
\begin{equation}
\label{pfinal}
p = \frac{d-3}{16\pi r \sqrt{1-\frac{r_+^{d-3}}{r^{d-3}}}}
\left( 1 - \sqrt{1-\frac{r_+^{d-3}}{r^{d-3}} }\right)^2 .
\end{equation}

To obtain the entropy $S$, we use Eqs.~(\ref{actionr+beta}) and
~(\ref{Efinal}) in Eq.~(\ref{S}) to yield
\begin{equation}\label{Sfinal}
S = \frac{\Omega_{d-2} r_+^{d-2}}{4}.
\end{equation}
This is the 
Bekenstein-Hawking entropy for black holes
in $d$ dimensions.

Having derived the important thermodynamic quantities,
we can now find how the number of dimensions $d$ affects 
the Euler relation and the Gibbs-Duhem
relation. From the equations for the thermodynamic energy and entropy,
Eqs.~(\ref{Efinal}) and (\ref{Sfinal}), we can write
$E=\frac{(d-2)\Omega_{d-2}^{\frac1{d-2}}}{8\pi} A^{\frac{d-3}{d-2}}
\left(1- \sqrt{1- \left( \frac{4S}{A} \right)^{\frac{d-3}{d-2}}} 
\right)$. So from Euler's theorem on homogeneous functions, we find
that
$E$ is homogeneous of degree $\frac{d-3}{d-2}$ in $S$ and $A$, 
i.e.,
$\frac{d-3}{d-2}E = \left( \frac{\partial E}{\partial S} \right)S
+ \left( \frac{\partial E}{\partial A} \right)A $, which means
\begin{equation}\label{Euler}
\frac{d-3}{d-2}E = TS - pA\,.
\end{equation}
This is the Euler relation for $d$-dimensional black holes in the
canonical ensemble.  Taking the differential of the Euler relation in
Eq.~(\ref{Euler}) and using the first law in Eq.~(\ref{firstlaw}), we
obtain
\begin{equation}\label{gibbs}
dE + (d-2)SdT -(d-2)A dp =0,
\end{equation}
which is the Gibbs-Duhem relation for $d$-dimensional
black holes.  In addition, the scaling laws for the gravitational
canonical ensemble in $d$ dimensions can be deduced to be $r
\rightarrow \lambda r$ ($A\rightarrow \lambda^{d-2}A$), $T\rightarrow
\lambda^{-1}T$ ($\beta\rightarrow \lambda \beta$), $E\rightarrow
\lambda^{d-3}E$, $S\rightarrow \lambda^{d-2}S$.  Curved space is
responsible for the fact that intensive parameters lose their
homogeneity of degree zero, i.e., the Tolman temperature formula for
thermal equilibrium in curved space forces the temperature to lose its
usual intensive character.  The same happens with the pressure, which
now scales as $p\rightarrow \lambda^{-1}p$, a scaling that comes about
because it is a pressure that acts in an area $A$ rather than in a
volume.  Consequently, extensive parameters such as the energy also
lose their homogeneity of degree 1.  The action $I$ scales as
$I\rightarrow \lambda^{d-2}I$, and the free energy $F$
scales as $F\rightarrow
\lambda^{d-3}F$.

\section{Thermal stability}
\label{secStability}

The heat capacity at constant cavity area, $C_A$, 
defined by
\begin{equation}
\label{cadef}
C_A \equiv \left( \frac{\partial E}{\partial T} \right)_A \,,
\end{equation}
determines the thermal stability
of a system in the canonical ensemble.
The thermodynamic energy $E(r_+,r)$ is given in Eq.~(\ref{E}),
and $r_+(\beta,r)$ is given through Eq.~(\ref{polynomial}).
Since
$T=\frac{1}{\beta}$,
see Eq.~(\ref{beta}), and since
$A={\rm const}$
implies $r={\rm const}$, see Eq.~(\ref{areacav}),
one finds that
$\left(
\frac{\partial E}{\partial T}
\right)_A 
 = - \beta^2
\frac{ \left( \partial E / \partial r_+ \right)_r}
{\left( \partial \beta / \partial r_+\right)_r}$.
Then, the heat capacity
for a black hole in $d$ dimensions is given by
\begin{equation}\label{cafinal}
C_A = \frac{(d-2)}{2(d-1)} \Omega_{d-2} r_+r^{d-3} 
\frac{
 1-\frac{r_+^{d-3}}{r^{d-3}} 
}
{
1 - \frac{2}{d-1} \frac{r^{d-3}}{r_+^{d-3}}
}.
\end{equation}
A system is thermally stable if 
\begin{equation}\label{cageq0}
C_A\geq0\,.
\end{equation}
Using Eq.~(\ref{cafinal}) on Eq.~(\ref{cageq0})
yields $r_+\leq r\leq (\frac{d-1}{2})^{\frac{1}{d-3}}r_+$.
Since $(\frac{d-1}{2})^{\frac{1}{d-3}}r_+$
is the photon orbit radius,  $r_{\rm ph}= \left(
\frac{d-1}{2} \right)^\frac1{d-3}r_+$, see also 
Appendix~\ref{phbuch},
one has
\begin{equation}\label{castability}
{r_+ \leq r \leq r_{\rm ph}}\,,
\end{equation}
i.e., the cavity's 
boundary $r$ must lie between the black hole and its photon
sphere radius, see Fig.~\ref{BHs}, and see also~\cite{GregRoss}.
The smaller
black hole $r_{+1}$
given in Eq.~(\ref{bh1}) will always have its photon
sphere inside the cavity radius $r$
and so is thermodynamically unstable.
The larger
black hole $r_{+2}$
given in Eq.~(\ref{bh2}) will have its photon
sphere outside the cavity radius $r$
and so is thermodynamically stable.

It is interesting to comment on the appearance of the photon orbit
radius, $r_{\rm ph}$, in the context of thermodynamics of black holes,
more precisely, in the context of black holes in the canonical
ensemble, see also Appendix~\ref{phbuch}.  The photon orbit radius
appears naturally in the context of particle dynamics in a
Schwarzschild background.  At this radius, massless particles
traveling at the speed of light can have circular orbits.  In four
dimensions, the photon orbit radius is $r_{\rm ph}=\frac32r_+$, in five
dimensions, it is $r_{\rm ph}=\sqrt2 r_+$, and in generic $d$
dimensions, it is $r_{\rm ph}= \left( \frac{d-1}{2}
\right)^\frac1{d-3}r_+$
\cite{monteiro}.  It is a surprise that the bound also appears
in a thermodynamic context. In this context, the bound states that in a
canonical ensemble with the boundary radius given by $r$, the black
hole is thermodynamically marginally stable if $r_{\rm ph}= r$, is
unstable if $r_{\rm ph}< r$, and stable if $r_{\rm ph}> r$.  The two
contexts, particle dynamics in a Schwarzschild background on one side
and black hole thermodynamic stability on the other, are somehow
correlated, although this correlation has not been clearly interpreted.


\section{Generalized free energy function}
\label{secFreeEnergyFunc}

Thermodynamics is valid for stationary and thermodynamic stable
systems.  We have seen that there are two black hole solutions.  One,
the small black hole solution with horizon radius $r_{+1}$, is
unstable, the other, the large black hole solution with horizon radius
$r_{+2}$, is stable.  So, the whole thermodynamic procedure is valid
in principle only for the $r_{+2}$ black hole.  For this black hole,
there is a well-defined action $I(r,r_{+2})$ given in
Eq.~(\ref{actionbh2}) in an approximation, and its thermodynamic free
energy is also well defined since $F(r,r_{+2}) =
\frac{I(r,r_{+2})}{\beta}$, see Eq.~(\ref{FandI}).

We can perturb the free energy $F$ by
keeping fixed the quantities that define the
canonical ensemble, precisely,
the cavity radius $r$ and temperature $T$, 
and allow $r_+$ to vary from zero ro $r$. 
This generalized free energy, $\bar{F}$, is then
\begin{align}
\bar{F}(r_+;r,T) =& \frac{(d-2)\Omega_{d-2} \, r^{d-3}}{8\pi}  
\times \nonumber \\
&\left( 1- \sqrt{1- \frac{r_+^{d-3}}{r^{d-3}}} - 
\frac{2\pi r T}{(d-2)} \left( \frac{r_+}{r} 
\right)^{d-2} \right)\,,
\label{Fbar}
\end{align}
valid for
$0\leq r_+\leq r$, and 
where we have used Eq.~(\ref{FandETS}) together with
Eqs.~(\ref{Efinal}) and (\ref{Sfinal}).

The generalized free energy $\bar{F}$ in Eq.~(\ref{Fbar}) has several
important properties.  For $r_+=0$, i.e., the situation where
there is no black hole, one has
$F=0$. The no black hole situation represents classical hot flat
space, i.e., nothing in a Minkowski spacetime, and so it is consistent
that it has zero free energy.  Also, $\bar{F}$ has two stationary
points as one readily finds by computing $\left( \frac{\partial
\bar{F}}{\,\partial r_+} \right)_{r,T}=0$.  The first stationary point
is a local maximum and can be seen to correspond
to the small black hole $r_{+1}$,
with $\bar{F}(r_{+1})$,
in thermal
equilibrium but unstable, see Eq.~(\ref{polynomial}).  The second
stationary point is a local minimum and can be seen to correspond to
the large black hole $r_{+2}$, with $\bar{F}(r_{+1})$, in
thermodynamic equilibrium and stable, see Eq.~(\ref{polynomial}).
Interpreting $\bar{F}$ as the thermodynamic potential of the ensemble,
one can say that the smaller black hole solution $r_{+1}$ acts as a
potential barrier separating two stable solutions, classical hot flat
space at $r_+=0$ with $\bar{F}=0$, and the large black hole $r_{+2}$
with $\bar{F}=\bar{F}(r_{+2})$. In general, $\bar{F}(r_{+2})\leq
\bar{F}(r_{+1})$,
i.e.,
$\pi r T \geq \frac{d-3}{4}\left[
\left( \frac{2}{d-1} \right)^{\frac{2}{d-3}}
-\left( \frac{2}{d-1} \right)^{\frac{d-1}{d-3}}\right]^{-1/2}$
the equality holding when $r_{+2}=r_{+1}$.

\begin{widetext}

\begin{figure}[t]
\centering
  \subfloat[\label{subfiga1}]{\includegraphics
  [width=.45\textwidth]{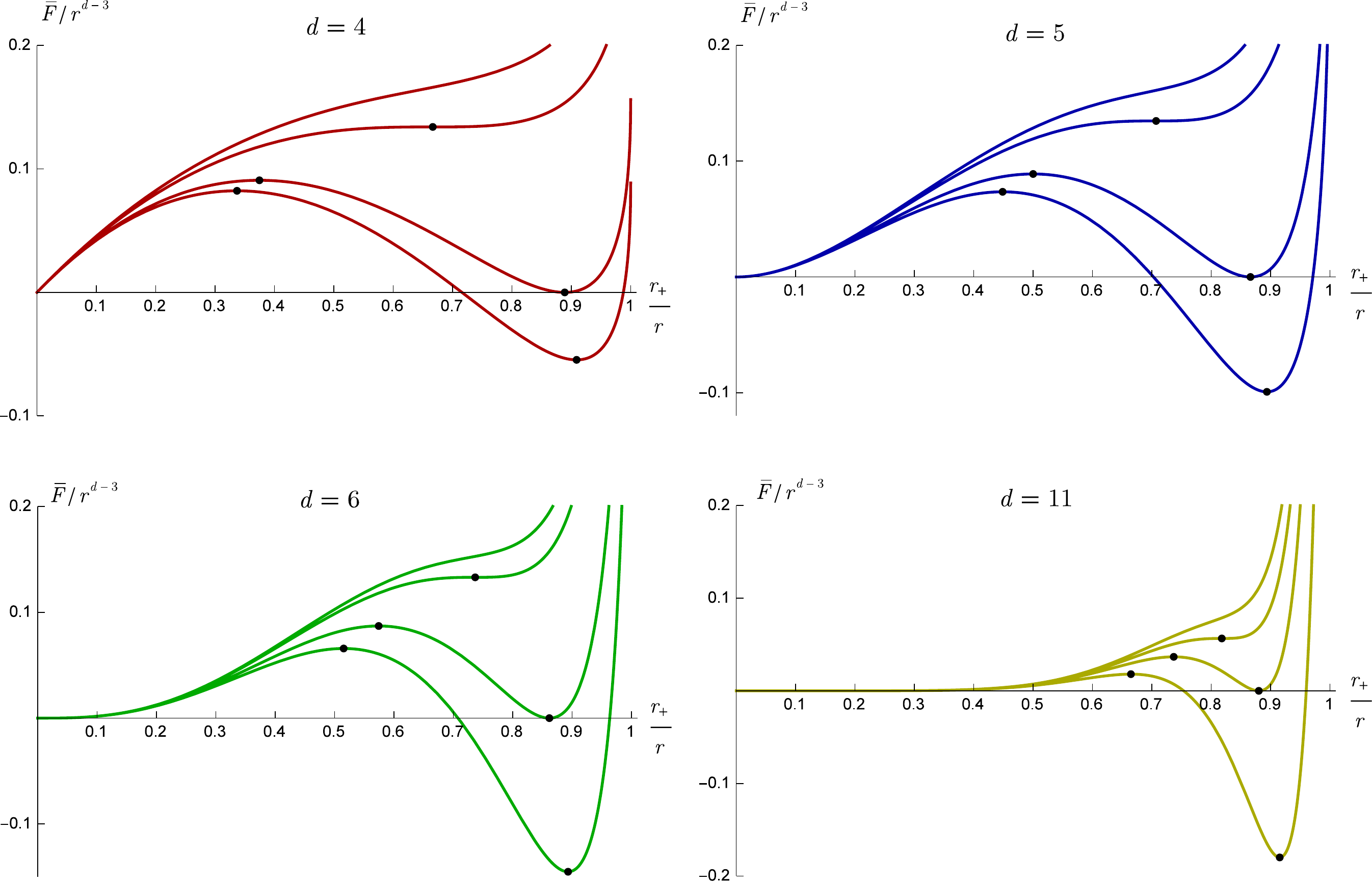}} 
  \hspace{5mm}
  \subfloat[\label{subfigb1}]
  {\includegraphics[width=.45\textwidth]{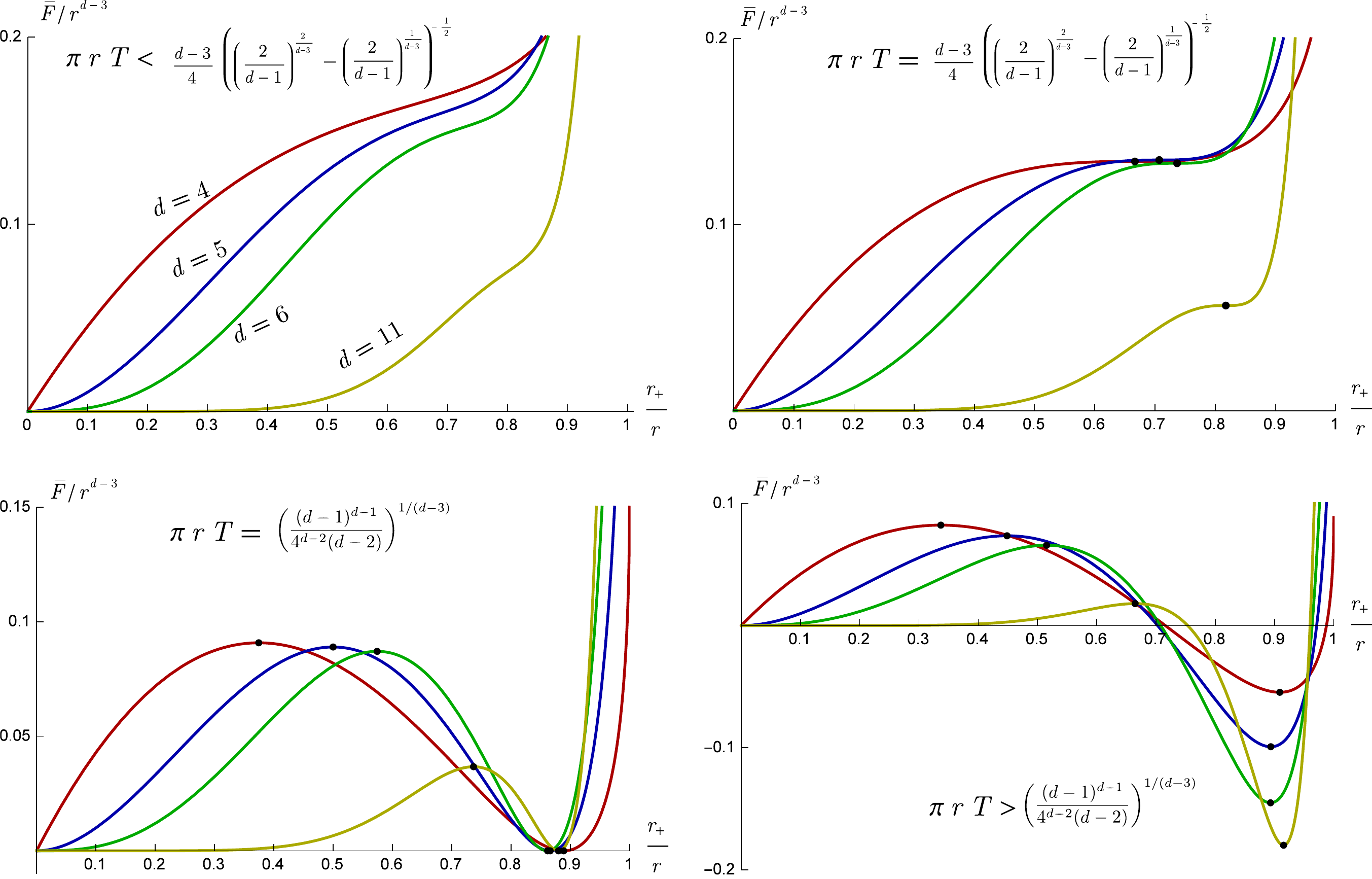}}
\caption{Parts (a) and (b) of the figure are
complementary, they contain the same information but present it
differently.
{\bf (a)}
The free energy function $\bar{F}$ is plotted as a function
of the horizon radius $r_+$ for four different dimensions, $d = 4$,
$5$, $6$, and $11$.  For each dimension, one has
the four typical curves
shown according to the value of $\pi r T$.  In each plot, the free
energy $\bar{F}$ is adimensionalized in terms of the cavity radius $r$
as $\frac{\bar{F}}{r^{d-3}}$, and the horizon radius is also normalized
to $r$ as $\frac{r_+}{r}$, so that $0\leq\frac{r_+}{r}\leq1$.  The
black hole solutions $r_{+1}$ and $r_{+2}$, when they exist, occur at
the extrema $\left( \frac{\partial {\bar F}}{\;\,\partial r_+}
\right)_{r,T} = 0$, the black dots serve to indicate them.  For each
dimension, the upper curve is for $\pi r T<\frac{d-3}{4}\left[ \left(
\frac{2}{d-1} \right)^{\frac{2}{d-3}} -\left( \frac{2}{d-1}
\right)^{\frac{d-1}{d-3}} \right]^{-\frac12}$, when there are no black
hole solutions.  For each dimension, the curve below the upper curve
is for the limiting situation $\pi r T=\frac{d-3}{4}\left[ \left(
\frac{2}{d-1} \right)^{\frac{2}{d-3}} -\left( \frac{2}{d-1}
\right)^{\frac{d-1}{d-3}} \right]^{-\frac12}$, where the two black
hole solutions coincide, $r_{+1}=r_{+2}$, in an inflection point, in
which situation there is neutral equilibrium.  For each dimension, the
curve above the lower curve is for $ \pi r T=
\left(\frac{(d-1)^{d-1}}{4^{d-2}(d-2)} \right)^{\frac{1}{d-3}}$, where
the smaller black hole $r_{+1}$ has positive free energy and is
unstable, and the larger black hole $r_{+2}$ has zero free energy and
is stable.  For each dimension, the lower curve is for $ \pi r T>
\left(\frac{(d-1)^{d-1}}{4^{d-2}(d-2)} \right)^{\frac{1}{d-3}}$, where
the smaller black hole $r_{+1}$ has still positive free energy and is
unstable, and the larger black hole $r_{+2}$ has now negative free
energy and is stable.  In the upper two curves, it is not possible for
classical hot flat space $r_+=0$ which has zero free energy to
transition to the large $r_{+2}$ black hole, but the $r_{+2}$ black
hole can transition to classical hot flat space.  In the lower two
curves, classical hot flat space $r_+=0$ can nucleate into the large
$r_{+2}$ black hole through the small black hole $r_{+1}$. 
{\bf (b)}
The free energy function $\bar{F}$ is plotted as a function
of the horizon radius $r_+$
for the four typical  situations, 
namely, $\pi r T<\frac{d-3}{4}\left[ \left( \frac{2}{d-1}
\right)^{\frac{2}{d-3}} -\left( \frac{2}{d-1}
\right)^{\frac{d-1}{d-3}} \right]^{-\frac12}$, $\pi r
T=\frac{d-3}{4}\left[ \left( \frac{2}{d-1} \right)^{\frac{2}{d-3}}
-\left( \frac{2}{d-1} \right)^{\frac{d-1}{d-3}} \right]^{-\frac12}$, $
\pi r T= \left(\frac{(d-1)^{d-1}}{4^{d-2}(d-2)}
\right)^{\frac{1}{d-3}}$, and $ \pi r T>
\left(\frac{(d-1)^{d-1}}{4^{d-2}(d-2)} \right)^{\frac{1}{d-3}}$. 
For each typical  situation, one has the four curves
corresponding to the four different dimensions,
$d=4$, $d=5$, $d=6$, and $d=11$.
In each plot, the free
energy $\bar{F}$ is adimensionalized in terms of the cavity radius $r$
as $\frac{\bar{F}}{r^{d-3}}$ and the horizon radius is also normalized
to $r$ as $\frac{r_+}{r}$, so that $0\leq\frac{r_+}{r}\leq1$.
}
\label{Fs}
\end{figure}

\end{widetext}

Moreover, $\bar{F}$ given in Eq.~(\ref{Fbar}) also signals phase
transitions. In the canonical ensemble, phase transitions occur always
in the direction of decreasing free energy, in this case decreasing
$\bar{F}$.
One can then study whether there is no possibility of the occurrence
of a phase transition from classical hot flat space to the stable
black hole $r_{+2}$ or, what here amounts to the same thing, whether
there is the possibility that a phase transition from the stable black
hole $r_{+2}$ to classical hot flat space can occur, and in which
conditions. One can also study, complementarily, whether there is the
possibility of the occurrence of a phase transition from classical hot
flat space to the stable black hole $r_{+2}$, and in which
conditions.
Figure~\ref{Fs} gathers all the necessary information
to study these phase transitions
by plotting the free
energy function $\bar{F}$ as a function of the horizon radius in units
of the cavity radius, $\frac{r_+}{r}$, as given in Eq.~(\ref{Fbar}),
for four different dimensions, $d=4$, $d=5$, $d=6$, and $d=11$, and
for each dimension, giving the four important different situations
that depend on the value of $\pi r T$, and, to complement, by also
plotting the free energy function $\bar{F}$ as a function of the
horizon radius in units of the cavity radius, $\frac{r_+}{r}$, as
given in Eq.~(\ref{Fbar}), for the four important different situations
that depend on the value of $\pi r T$, and in each situation showing
the four different dimensions, $d=4$, $d=5$, $d=6$, and $d=11$.
Let us see in detail these phase transitions. 
We start the analysis in the case that there is no possibility of the
occurrence of a phase transition from classical hot flat space to the
stable black hole $r_{+2}$. Or, what here amounts to the same thing, in
the case that a phase transition from the stable black hole $r_{+2}$ to
classical hot flat space can occur.
Since classical hot flat space has zero free
energy
$\bar{F}=0$, one has that a phase transition
from the stable black hole $r_{+2}$
to classical hot flat space can occur
when $\bar{F}(r_{+2})\geq0$, i.e., 
$I(r_{+2})\geq0$.
By repeating the analysis
done from Eq.~(\ref{actionr+})
to
Eq.~(\ref{buch}), one can find 
that $I(r_{+2})\geq0$ when
$\frac{r_+}{r} \leq
\frac{r_+}{r_{\rm Buch}}$.
where $r_{\rm Buch}$ is the $d$-dimensional
Buchdahl radius
given by 
$r_{\rm Buch}=\left( \frac{ (d-1)^2 }{ 4(d-2) }
\right)^{\frac{1}{d-3}}r_+$.
Together with the condition for the existence of
black holes in thermodynamic equilibrium,
i.e., Eq.~(\ref{bhcondition}),
one finds 
that a large black hole $r_{+2}$ can decay into
classical hot flat space when
\begin{align}\label{fullc}
&\frac{d-3}{4}\left[
\left( \frac{2}{d-1} \right)^{\frac{2}{d-3}}
-\left( \frac{2}{d-1} \right)^{\frac{d-1}{d-3}}
\right]^{-\frac12}
\nonumber \\
&\leq\pi r T\leq
\left(\frac{(d-1)^{d-1}}{4^{d-2}(d-2)}
\right)^{\frac{1}{d-3}}\,,\quad
r_{\rm Buch}\leq r<\infty\,.
\end{align}
Also, when $\pi r T$ and $r$ obey
Eq.~(\ref{fullc}),
classical hot flat space never nucleates into a black hole.
We now analyze the inverse transition, i.e.,
the transition from classical hot flat
space to the stable black hole.
Since classical hot flat space has zero free energy
$\bar{F}=0$, one has that a phase transition
to the stable black hole $r_{+2}$ can occur
when $\bar{F}(r_{+2})\leq0$, i.e., 
$I(r_{+2})\leq0$.
From Eq.~(\ref{actionr+}),
we have done the analysis ending in 
Eq.~(\ref{buch}), i.e.,
we have found that $I(r_{+2})\leq0$ when
$\frac{r_+}{r} \geq
\frac{r_+}{r_{\rm Buch}}$
where $r_{\rm Buch}$ is the $d$-dimensional
Buchdahl radius
given by 
$r_{\rm Buch}=\left( \frac{ (d-1)^2 }{ 4(d-2) }
\right)^{\frac{1}{d-3}}r_+$.
Putting this back into Eq.~(\ref{Fbar}),
one can see that this happens
for $\pi r T \geq \left( 
\frac{(d-1)^{d-1}}{4^{d-2}(d-2)} \right)^{\frac{1}{d-3}}$.
Thus, a transition from classical hot flat
space to the stable black hole
occurs for 
\begin{equation}\label{F<0}
\left( 
\frac{(d-1)^{d-1}}{4^{d-2}(d-2)} \right)^{\frac{1}{d-3}}
\leq
\pi r T \leq \infty
\,,\quad
r_+\leq r\leq 
r_{\rm Buch}
\,.
\end{equation}
Equation~(\ref{F<0})
is a necessary and sufficient condition for
the occurrence of nucleation
from classical hot flat space to the stable black hole $r_{+2}$,
a transition that is done 
through the unstable black hole $r_{+1}$.
We also see that 
Eq.~(\ref{F<0}) imposes a
stronger condition than the
$\pi r T \geq \frac{d-3}{4}\left[
\left( \frac{2}{d-1} \right)^{\frac{2}{d-3}}
-\left( \frac{2}{d-1} \right)^{\frac{d-1}{d-3}}\right]^{-\frac12}$
of Eq.~(\ref{bhcondition})
for having black holes  in thermodynamic equilibrium at all.


It is interesting to comment on the appearance of the Buchdahl
radius, $r_{\rm Buch}$, in the context of thermodynamics of black
holes, more precisely, in the context of black holes in the canonical
ensemble, see also Appendix~\ref{phbuch}.  The Buchdahl bound has
appeared in the context of general relativistic star structure.  It is
a bound that states that under some generic conditions for a star of
radius $r$, the spacetime is free of singularities for $r_{\rm
Buch}\leq r$.  It is a lower bound for the ratio $\frac{r}{r_+}$,
where $r$ is the star's radius and $r_+$ its gravitational radius,
that appears such that the star spacetime is singularity free.
Presumably, for $r_{\rm Buch}\geq r$, the star might collapse into a
black hole.  In four dimensions, the limiting radius of the bound is
$r_{\rm Buch}=\frac98r_+$, in five dimensions, it is $r_{\rm
Buch}=\frac2{\sqrt3} r_+$, and in generic $d$ dimensions, the limiting
radius of the bound is $r_{\rm Buch}=\left( \frac{ (d-1)^2 }{ 4(d-2) }
\right)^{\frac{1}{d-3}}r_+$, see \cite{buchdahl,andreasson} for four
dimensions and \cite{wright} for $d$ dimensions.  It is a surprise
that the bound also appears in a thermodynamic context. 
In this
context, the bound we have found
states that in a canonical ensemble with the
boundary radius given by $r$,
classical hot flat space cannot 
transition to a black hole phase if
$r_{\rm Buch}\leq r$.  If, contrarily, $r_{\rm Buch}\geq r$, then
classical hot
flat space can make a transition to a black hole.
The two contexts,
general relativistic star solutions and gravitational collapse on one
side and black hole thermodynamic on the other, are thus clearly
correlated, and and thus this
correlation hints that $r_{\rm Buch}$ is an intrinsic property
of the Schwarzschild spacetime, as the radius of the photon orbit,
$r_{\rm ph}$, is.
To corroborate this statement and explicitly
see
this correlation, a comparison of
the thermodynamics of Schwarzschild black
holes and classical hot flat
space in a cavity with radius $r$ at a fixed temperature $T$ in
the canonical ensemble in $d$ dimensions
with the thermodynamics of a self-gravitating
thin shell of radius $r$ and at temperature $T$
with a Minkowski interior and
a Schwarzschild exterior can be performed, 
see Appendix~\ref{secshells}.


\section{Action functional to second order and its role in
thermodynamic stability}
\label{sa}

The path integral approach to a quantum gravity system prescribes that
one must integrate the
the exponential of the negative
of the Euclidean Einstein action $I$ over the space of
metrics $\rm g$ to obtain the canonical partition function of the
system, $Z= \int d{\rm [g]} \, \exp(-I[\rm g])$.
For a black hole
system with classical action $I$, one can use
the zeroth order
approximation yielding $Z= \exp(-I)$, see
Eq.~(\ref{partition}).  One can go a step further and perturb the
Euclidean black hole metric ${\bar g}_{ab}$ by a small amount
$h_{ab}$, such that the full perturbed metric is $g_{ab}={\bar
g}_{ab}+h_{ab}$, where clearly
${\bar g}_{ab}$ is envisaged now as a
background solution and $h_{ab}$ is envisaged
as a small fluctuation.
The Euclidean action can
then be approximated by
$I[{\rm g}] = I[\bar{{\rm g}}] + \int{d^4x \sqrt{\bar{\rm g}}\,
A_{abcd}h^{ab}h^{cd}}$,
for some operator $A_{abcd}$
which generically depends on the metric ${\bar g}_{ab}$,
its covariant derivatives, and curvature terms.
There are two possibilities depending on the perturbation operator
$A_{abcd}$.
If one of the eigenvalues of $A_{abcd}$ is negative,
then the integral gets an imaginary term, which implies
that the action and the free energy have an imaginary term
and the partition
function will also contain an imaginary part. 
In this case, the original classical black hole instanton is
a saddle point, and it is unstable.
On the contrary, there is the possibility
that all of the eigenvalues of $A_{abcd}$ are positive, in
which case
the perturbation modes are stable around the given black hole
solution.

In four dimensions,
the perturbation performed 
around the Euclidean Schwarzschild black hole
solution with a cavity with a very large
radius $r$ at a fixed temperature $T$
yielded  
that the
operator $A_{abcd}$ has indeed a negative eigenvalue,
resulting in an instability \cite{gpy}.
Connecting this result to thermodynamic
stability, it means that
a black hole in thermodynamic equilibrium in the canonical
ensemble with a large cavity cannot be thermodynamically stable.
However, when the cavity radius $r$ is reduced,
one finds \cite{allen} that 
the negative mode vanishes below a
certain radius $r=\frac32r_+$ of the cavity, indicating
stability.
Connecting the result to thermodynamic
stability, it means that
a black hole in thermodynamic equilibrium in the canonical
ensemble can be thermodynamically stable for $r\leq\frac32r_+$.
This correspondence between perturbation
path integral theory and thermodynamic stability was
found by York 
\cite{york1}, 
establishing that there
are actually two Euclidean solutions in thermal
equilibrium,
one of which is in an unstable thermodynamic
equilibrium which has negative
heat capacity, the smaller one  denoted by
$r_{+1}$,
and one
which is in stable thermodynamic
equilibrium which has positive
heat capacity, the larger one 
denoted by
$r_{+2}$.
The condition for stable thermodynamic equilibrium matched
exactly the condition for stability of the solution.

In $d$ dimensions, one can also work out a perturbation analysis on the
path integral \cite{GregRoss} to find that $d$-dimensional
Schwarzschild black holes have a negative mode if the black hole
radius $r_+$ is small compared to the cavity radius $r$, i.e., there
is a negative mode for the $r_{+1}$ black hole, and have no negative
mode if the black hole radius $r_+$ is of the order of the cavity
radius $r$, i.e., there is no negative mode for the $r_{+2}$ black
hole, the marginal zero mode case being when the cavity radius $r$ is
at the photon orbit radius $r=r_{\rm ph}$.  These results are thus
also in one-to-one correspondence with the instability or stability
thermodynamic analysis done through the heat capacity of the
$d$-dimensional black hole.  In~\cite{reallbranes}, it was further
clarified that thermodynamic stability of black holes and the mechanic
stability of black systems, such as black branes, are interrelated.

\section{Ground state of the canonical ensemble: quantum hot
flat space, black hole, or both}
\label{secGround}

From the 
partition function of black holes and the thermodynamic stability,
as well as from the perturbation studies on the action functional,
it is clear that in order to
properly understand the physics involved, one has to treat hot flat
space in quantum terms, i.e., hot flat space
should be treated as made of hot gravitons.
In this way, the issues of what is the ground state of the canonical
ensemble and what are the possible phase transitions can be addressed.

In the canonical ensemble, the ground state is the one that has
the lowest free energy $F$ or, if one prefers, the lowest
action $I$, as $I=\beta F$.
For the hot gravity system
under study, the three possible phases are 
quantum hot flat space,
the phase of a stable black hole with large radius $r_{+2}$,
or a possible superposition of these two  phases.
The black hole with small radius $r_{+1}$ is not a phase
since it is  unstable, as found previously.
Thus, to find the ground state
of hot gravity in the canonical ensemble, i.e.,
hot gravity at a given temperature $T$
and a given cavity radius $r$, the free energy of
quantum hot flat space $F_{\rm HFS}$ and
the free energy of the
large black hole $F(r_{+2})$ must be compared.

Minkowski flat space has $r_+ = 0$
and in the context of hot gravity is also
a solution in thermal equilibrium, 
i.e., fixing 
the temperature at the cavity boundary, the temperature will be
the same everywhere inside the cavity.
As follows from the 
Stefan-Boltzmann law, quantum hot flat space, or
flat space at finite temperature,
has finite free energy and thus finite action.
For a $d$-dimensional system containing only gravitons,
which is the case we consider here, 
the number of massless species is given by 
$\frac12 d(d-3)$, and in this case, one finds 
that the free energy in $d$ dimensions
of quantum hot flat space is given by
\begin{equation}\label{achfs} 
F_{\rm HFS} = - \frac{d(d-3)}{2(d-1)^2} a\, \Omega_{d-2}\, r^{d-1} T^d\,,
\end{equation}
where
$a = \frac{\Gamma(d)\zeta(d)}{2^{d-2}\pi^{\frac{d-1}{2}}\Gamma 
\left( \frac {d-1}2 \right)}$, with $\Gamma$ and
$\zeta$ being the gamma and zeta functions, respectively, see 
Appendix~\ref{secApB}.
The free energy of quantum hot flat space is negative,
and not zero as in the case of classical hot
flat space. Its
dependence with the cavity radius $r$ and temperature
$T$ is $r^{d-1} T^d$.
If one prefers to use the action $I$, 
then since $I=\beta F$ and $\beta=\frac1T$
one has
$I_{\rm HFS} = - \frac{d(d-3)}{2(d-1)^2}\,
a\, \Omega_{d-2}\, r^{d-1} T^{d-1}$.

The free energy of the stable black hole $r_{+2}$ is
$F(r_{+2})$. This free energy can be found
using the larger $r_{+2}$ solution of Eq.~(\ref{polynomial})
in Eq.~(\ref{actionr+beta}) for $I$ and then using 
$F(r_{+2})=TI(r_{+2})$.
For $d=5$, one can find
an exact solution \cite{andrelemosd5}, but
for any other $d$, either there is no exact solution
or, if there is, it is unusable.
We can then either resort to the large $T$ approximation
for $r_{+2}$ given in Eq.~(\ref{bh2})
and for $I(r_{+2})$ and so $F(r_{+2})$ given in 
Eq.~(\ref{actionbh2}) or to numeric calculations.
Let us start with the  large $T$ approximation.
Using Eq.~(\ref{actionbh2})
and $F(r_{+2})=TI(r_{+2})$ yields
\begin{align}\label{fr+2} 
F(r_{+2}) &= - \frac{\Omega_{d-2} r^{d-2}T}{4}\left( 
1 - \frac{d-2}{2 \pi r T} + 
\frac{(d-2)(d-3)}{16\left(\pi r T\right)^2} \right) +
\\ \nonumber
&+\mathcal O \left( 
\frac{d^4}{\left( \pi r T \right)^{4}}\right),  
\end{align}
where  we have been shortening the notation
$F(r_{+2}) \equiv F(r,r_{+2}(r,\beta))$.
One could make a plot through numerical calculations
of
$F(r_{+2})$ as a function of
$\pi r T$, but it is not so useful.

\begin{figure}[t]
\centering
\includegraphics[width=0.44\textwidth]{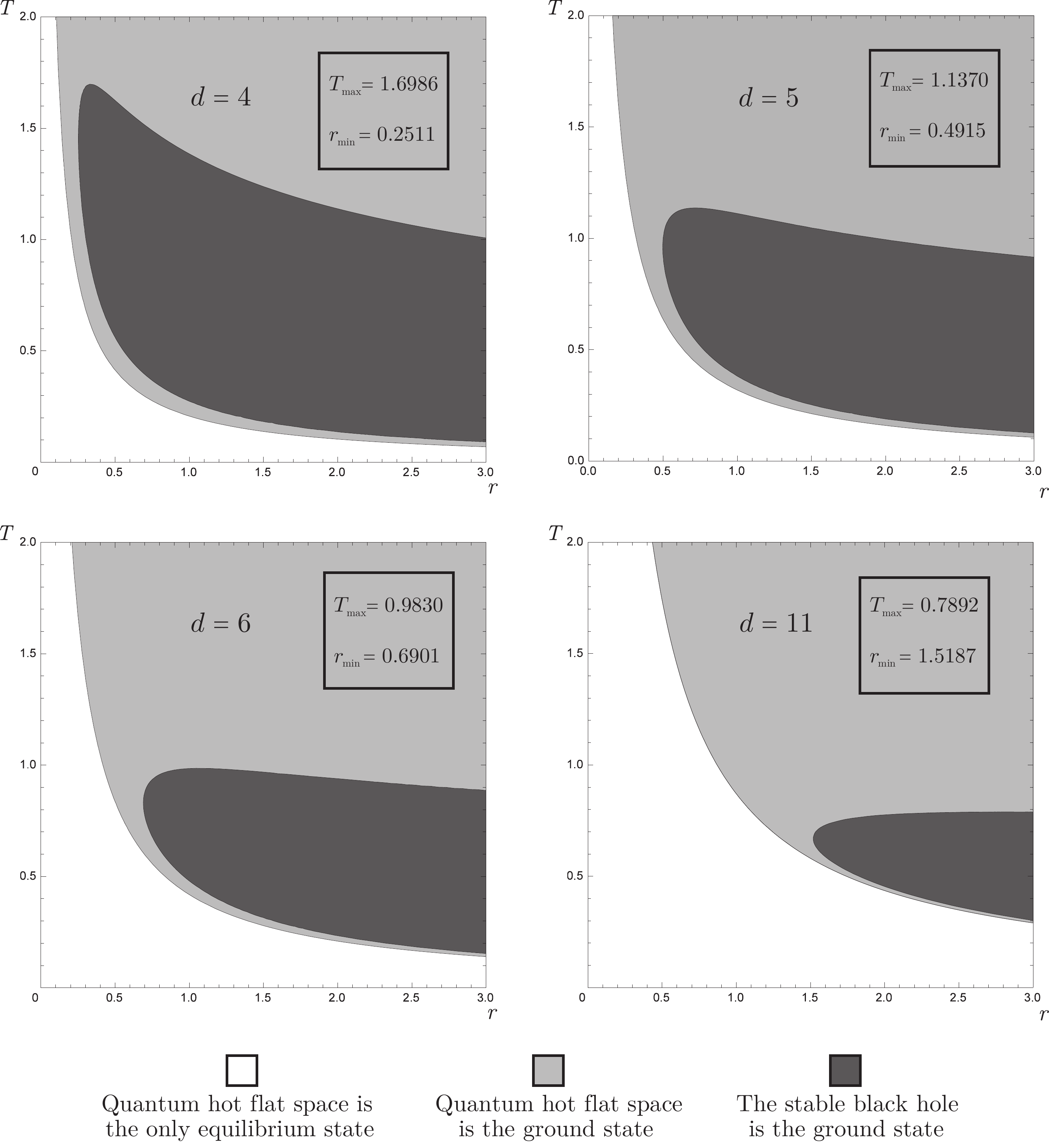}
\caption{
Phase diagram of the gravitational canonical ensemble with a plot of
the cavity radius $r$ versus the temperature $T$, both in Planck
units, for four different dimensions, $d=4$, $5$, $6$, and $11$.  Each
point in the plots represents a different configuration of the
ensemble.  Each plot is separated in three different regions that
yield three different phases.  The white region in each plot, for
which $\pi r T \geq \frac{d-3}{4}\left[ \left( \frac{2}{d-1}
\right)^{\frac{2}{d-3}} -\left( \frac{2}{d-1}
\right)^{\frac{d-1}{d-3}}\right]^{-1/2}$ holds, see
Eq.~(\ref{bhcondition}), is the region, or phase, where there quantum
hot flat space is the only equilibrium state, in this region there are
no black holes in thermodynamic equilibrium at all.  The gray region in
each plot characterizes the phase that has quantum hot flat space as
the ground state, i.e., the action for quantum hot flat space is lower
than the stable black hole's action, see Eq.~(\ref{fr+2versushfs})
with the inequality reversed.  The dark region in each plot
characterizes the phase that has the larger stable black hole $r_{+2}$
as the ground state of the canonical ensemble, see
Eq.~(\ref{fr+2versushfs}) with the inequality holding.  A quantum hot
flat space configuration in this phase is then able to nucleate stable
black holes.  The thick black line in each plot yields a mixed phase,
i.e., a superposition of the quantum hot flat ground state phase with
the stable black hole ground state phase, see
Eq.~(\ref{fr+2versushfs}) with the equality holding.  The extremal
values for the radius, $r_\text{min}$, and for the temperature,
$T_\text{max}$, are the lower bound and the upper bound of the dark
region, respectively.  If instead of quantum hot flat space one were
to consider classical hot flat space, i.e., the zero free energy of
Minkowski spacetime, the gray and the dark regions would be separated
by an asymptote following Eq.~(\ref{F<0}), which in the classical
limit, i.e., $ r\gg1$ and $T\ll1$, in Planck units, matches the line
separating both regions, see Appendix~\ref{appchfs} for further
details on classical hot flat space with the corresponding black hole
phase transitions.
}
\label{GroundStates}
\end{figure}

One has now to compare $F_{\rm HFS}$ of Eq.~(\ref{achfs})
with $F(r_{+2})$ of Eq.~(\ref{fr+2}) or $F(r_{+2})$
given by numerical calculations.
The stable black
hole $r_{+2}$ is the ground state when 
\begin{equation}\label{fr+2versushfs} 
F(r_{+2}) \leq F_{\rm HFS}\,.
\end{equation}
In the situation that the equality holds then
the black hole and quantum hot flat space phases coexist.
For the phase diagram of the gravitational canonical ensemble with a
plot of the cavity radius $r$ versus the temperature $T$ for several
different dimensions, specifically, $d=4$, $5$, $6$, and $11$, see
Fig.~\ref{GroundStates}.

Let us first use the
approximation given by
Eq.~(\ref{fr+2}). Putting Eqs.~(\ref{achfs}) and~(\ref{fr+2})
into Eq.~(\ref{fr+2versushfs}) yields 
\begin{align}
r^{-(d-2)} \leq& \frac{\pi^{d-1}(d-1)^2}{2 a d(d-3) (\pi r T)^{d-1}}
\times \nonumber \\
&\times \left(1-\frac{d-2}{2\pi r T}+\frac{(d-2)(d-3)}{16 (\pi r T)^2}
 \right),
\label{rediusIneq}
\end{align}
up to $\mathcal{O}\left(\frac1{(rT)^4}\right)$.
Note that the right-hand side of Eq.~(\ref{rediusIneq})
has an extremum at 
$\pi r T = \frac{d(d-2)+\sqrt{(d-2)(d^2+d-3)}}{4(d-1)}$, leading to a 
minimum radius $r_\text{min}$ given by
$r_\text{min} = \left(\frac{a d (d-3)}{2 \pi^{d-1}} \right)^\frac{1}{d-2} 
\times\left(\frac{\left( \frac{d(d-2) + 
\sqrt{(d-2)(d^2+d-3)}}{d-1} \right)^{d+1}}{2^{2d-3}(d-2)\left(2d-3 + 
\sqrt{(d-2)(d^2+d-3)}\right)} \right)^\frac{1}{d-2}$,
below which the black hole will never be the ground
state of the ensemble. 
In more detail,
in the case the radius of the cavity
is smaller than $r_\text{min}$
one has that quantum hot flat space is always the ground state.
Therefore, a necessary but not sufficient condition for black hole
nucleation from quantum hot flat space is that the
radius of the cavity be
greater than $r_\text{min}$.
One finds in these approximations
that
$r_\text{min}\simeq 0.2525$ in $d=4$, 
$r_\text{min}\simeq 0.4971$ in $d=5$, 
$r_\text{min}\simeq 0.7012$ in $d=6$, and
$r_\text{min}\simeq 1.5636$ in $d=11$.
In the large $d$ limit one has $r_\text{min}
\rightarrow a^{\frac1d}$,
and since
$a = \frac{\Gamma(d)\zeta(d)}{2^{d-2}\pi^{\frac{d-1}{2}}\Gamma 
\left( \frac {d-1}2 \right)}
\to\infty$, in the $d\to\infty$ limit, one has that $r_\text{min}$ tends to
infinity. For $d$ finite, say $d=11$,
$r_\text{min}\simeq 1.5636$, i.e., $r_\text{min}$ is still near the Planck
length which we have set to 1, but it increases for larger $d$.
If one uses numerical calculations,
see Fig.~\ref{GroundStates}, then one finds
$r_\text{min}\simeq 0.2511$ in $d=4$, 
$r_\text{min}\simeq 0.4915$ in $d=5$, 
$r_\text{min}\simeq 0.6901$ in $d=6$, and
$r_\text{min}\simeq 1.5187$ in $d=11$.
The approximation is in any case excellent.

The inequality of Eq.~(\ref{fr+2versushfs}) can also be written in
terms of the temperature of the cavity $T$.  Let us use the
approximation.  Then, indeed, Eq.~(\ref{fr+2versushfs}) yields
\begin{equation}
T^{d-2} \leq \frac{\pi(d-1)^2}{2 a d(d-3) \, \pi r T} 
\left( 1- \frac{d-2}{2 \pi r T} + 
\frac{(d-2)(d-3)}{16 (\pi r T)^2} \right),
\label{tempIneq}
\end{equation}
up to $\mathcal{O}\left(\frac1{(rT)^4}\right)$.
Now, the right-hand side has
an extremum at $\pi r T = \frac{d-2}{2}+\frac{\sqrt{(d-2)(d+1)}}{4}$
which then
leads to finding a maximum value for the temperature $T_\text{max}$ given by
$T_\text{max} = \left( \frac{4\pi}{a d(d-3)} \right)^{\frac1{d-2}}\times \left(
\frac{(d-2)(d-1)^2\left(d-1+\sqrt{(d-2)(d+1)}\right)}
{\left(2(d-2)+\sqrt{(d-2)(d+1)}\right)^3} \right)^{\frac1{d-2}}$,
above which the black hole will never be the ground state. 
In more detail, 
in the case the temperature is larger than $T_\text{max}$ one has that
quantum hot flat space is always the ground state.
Therefore, a necessary but not sufficient condition for black hole
nucleation from quantum
hot flat space is that the temperature of the cavity be
smaller than $T_\text{max}$.
One
finds in these approximations that
$T_\text{max}\simeq 1.6979$ in $d=4$, 
$T_\text{max}\simeq 1.1365$ in $d=5$, 
$T_\text{max}\simeq 0.9827$ in $d=6$, and
$T_\text{max}\simeq 0.7891$ in $d=11$.
In the large $d$ limit, one finds that $T_\text{max}$ tends to
$T_\text{max} \rightarrow
\frac{1}{a^{\frac1d}}$ which, taking into account the expression for
$a$, $a = \frac{\Gamma(d)\zeta(d)}{2^{d-2}\pi^{\frac{d-1}{2}}\Gamma
\left( \frac {d-1}2 \right)}$, given the $d\to\infty$ limit of $a$,
tends to $0$. For not so large values of $d$, the maximum temperature
does not deviate much from the Planck temperature which we set to 1.
If one uses numerical calculations, see Fig.~\ref{GroundStates}, then
one finds
$T_\text{max}\simeq 1.6986$ in $d=4$, 
$T_\text{max}\simeq 1.1370$ in $d=5$, 
$T_\text{max}\simeq 0.9830$ in $d=6$, and
$T_\text{max}\simeq 0.7892$ in $d=11$.
One sees that the approximations for the maximum temperature hold
better than for the minimum radius for large $d$.  The fact that the
approximation for $T_\text{max}$ holds better than the one for
$r_\text{min}$ for higher dimensions is because the initial
approximation taken, i.e., the solution $r_{+2}$ in Eq.~(\ref{bh2}),
depends on $\pi r T$. From Eq.~(\ref{rediusIneq}), it was seen that,
for large $d$, $r_\text{min}$ will lie on a curve approaching $\pi r T
= \dfrac{d}{4}$, whereas from Eq.~(\ref{tempIneq}), for large $d$,
$T_\text{max}$ will lie on a curve approaching $\pi r T =
\dfrac{3d}{4}$, so $T_\text{max}$ will be more accurate than
$r_\text{min}$, since Eq.~(\ref{bh2}) holds better for larger values
of $\pi r T$.

Thus, from Fig.~\ref{GroundStates} we see that there are three phases.
One phase is when the cavity's radius $r$ and the cavity's temperature
$T$ are such that in
thermodynamic equilibrium only quantum hot flat space is
possible, there are no stable equilibrium black holes $r_{+2}$, and
for that matter there are
also no unstable equilibrium black holes $r_{+1}$, but
black holes out of thermodynamic equilibrium may perhaps appear in
this phase. Another phase when the cavity's radius $r$ and the
cavity's temperature $T$ are such that
quantum hot flat space is the ground state and so
stable black holes $r_{+2}$ can
transition into quantum hot flat space. And yet another phase is when the
cavity's radius $r$ and the cavity's temperature $T$ are such that
the stable black hole $r_{+2}$ is the ground state, and so 
quantum hot flat space can nucleate stable black holes.  These
three phases are represented by the white, gray, and dark regions,
respectively in
Fig.~\ref{GroundStates}. There is a mixed phase which
is a superposition
of the quantum hot flat ground state phase
with the 
stable black hole $r_{+2}$ ground state phase,
which is represented by a line between the gray
and dark regions. A feature that Fig.~\ref{GroundStates} makes clear
is that as the number of spacetime dimensions increases, the region for
the
quantum hot flat ground state phase
gets larger, whereas the region for the
stable black hole ground state phase
gets smaller.  In the $d\to\infty$ limit
black holes never nucleate as expected.

It is also of interest to understand the passage from quantum hot flat
space and black hole phase transitions to classical hot flat space and
the corresponding black hole phase transitions.  In this passage one
puts the constant $a$ that appears in Eq.~(\ref{achfs}) to zero,
$a=0$, and the analysis follows, see Appendix~\ref{appchfs}.

\section{Density of states}
\label{secDensity}

It is interesting to find through the density
of states $\nu$ with a given energy $E$
that the  entropy of the $r_{+2}$ black hole is 
$S=\frac{A_{+2}}{4}$.

Fixing the cavity radius $r$, the number of states between
$E$ and $E+dE$ is given by $\nu(E)dE$ with $\nu(E)$ 
being the density of states. 
Thus, weighing this density $\nu(E)$
with the Boltzmann factor
$e^{-\beta E}$, the
canonical partition function can be written as
$Z(\beta,r)= \int dE\, \nu(E)e^{-\beta E}$.
Inverting this expression
by an inverse Laplace transform 
one obtains
$\nu(E)= \frac1{2\pi i} \int_{-i \infty}^{i \infty} 
d\beta\, Z(\beta)e^{\beta E}$.
The partition function 
for the stable black hole is  $Z=\exp(-I(r,r_{+2}(r,\beta)))$.
Using $I(r,r_{+2}(r,\beta))$ given in 
Eq.~(\ref{actionbh2}) for large $rT$,
one finds
\begin{align}
Z(\beta,r) \simeq \exp& \left( \frac{\Omega_{d-2} r^{d-2}}{4} - 
\frac{(d-2)\Omega_{d-2} r^{d-3}}{8\pi}\beta + \right.  \nonumber \\
+&
\left.\frac{(d-2)(d-3)\Omega_{d-2} r^{d-4}}{64\pi^2} \beta^2 \right).
\end{align}
Taking the inverse Laplace transform, one has
$\nu(E) = \frac{4\sqrt\pi}{\sqrt{(d-2)(d-3)\Omega_{d-2} r^{d-4}}}
\exp\Big( \frac{\Omega_{d-2} r^{d-2}}{4} - 
\frac{16\pi^2}{(d-2)(d-3)\Omega_{d-2} r^{d-4}}\left(E - 
\frac{(d-2)\Omega_{d-2} r^{d-3}}{8\pi} \right)^2 \Big).$
Now, the spacetime mass $m$ 
is given in terms of $E$ and the cavity radius $r$ by
$m = E - \frac{4\pi E^2}{(d-2)\Omega_{d-2} r^{d-3}}$.
We can then write $\nu(E)$ as
$\nu(E) = \frac{4\sqrt\pi}{\sqrt{(d-2)(d-3)\Omega_{d-2} r^{d-4}}}
\exp\left( - \frac{\Omega_{d-2} r^{d-2}}{4(d-3)} +
\frac{4\pi r m}{(d-3)} \right)$.
Finally, with the equation
$r_+^{d-3}=\frac{16\pi}{(d-2)\Omega_{d-2}}\,m$ along with the fact that
$r_{+2}\simeq r$ for the stable black hole in this regime,
one finds that the density of states
is well described by
\begin{equation}
\nu(E) = \frac{4\sqrt{\pi}}
{\sqrt{(d-2)(d-3)\Omega_{d-2} r^{d-4}}}
\exp{\left( \frac{A_{+2}}{4} \right)},
\end{equation}
where the area of the black hole is
$A_{+2}= \Omega_{d-2} r_{+2}^{d-2}$.
The entropy $S$  and
the density of states $\nu$ are related through the
formula $S = a \ln \nu$,
for some constant $a$, so 
the black hole entropy is 
\begin{equation}
S=\frac{A_{+2}}{4}\,,
\end{equation}
where we discarded the remaining constant. 
In contrast, for the unstable black hole $r_{+1}$,
the action of Eq.~(\ref{actionbh1})
has a divergent integral when one performs
the Laplace transform.
Only the
large stable
black hole $r_{+2}$ yields the correct result.

For a synopsis of all the results and further comments see Appendix~\ref{conc}.

\section*{Acknowledgments}
RA acknowledges support from the Doctoral Programme in the Physics
and Mathematics of Information (DP-PMI) and the Funda\c c\~ao para a
Ci\^encia e Tecnologia (FCT)  through Grant
No.~PD/BD/135011/2017.  JPSL acknowledges FCT for financial
support through Project~No.~UIDB/00099/2020.

\vskip 1cm

%
%

\appendix
\label{appall}

\renewcommand\thefigure{\thesection\arabic{figure}}


\setcounter{figure}{0}

\section{Calculation of the approximate
expressions for the canonical ensemble
horizon radii $r_{+1}$
and $r_{+2}$}
\label{calculations}

Here we perform the calculation that lead to
the approximate expressions for $r_{+1}$
and $r_{+2}$ from
Eq.~(\ref{polynomial})
to Eqs.~(\ref{bh1}) and ~(\ref{bh2}),
respectively.
For the sake of
quick reference, we repeat
Eq.~(\ref{polynomial}), which is
\begin{equation}\label{polynomial1}
\left(\frac{r_+}{r} \right)^{d-1}-\left(\frac{r_+}{r} \right)^2+
\left( \frac{d-3}{4\pi r T} \right)^2 = 0\,.
\end{equation}
It is a polynomial equation of order
$d-1$, which has direct exact solutions for $d=4$ and $d=5$,
whereas for other $d$
one is compelled to resort to approximation
schemes or numerical calculations.
We display an approximation scheme
to find 
$r_{+1}$
and $r_{+2}$.

\begin{figure}[h]
\centering
\includegraphics[width=0.20\textwidth]{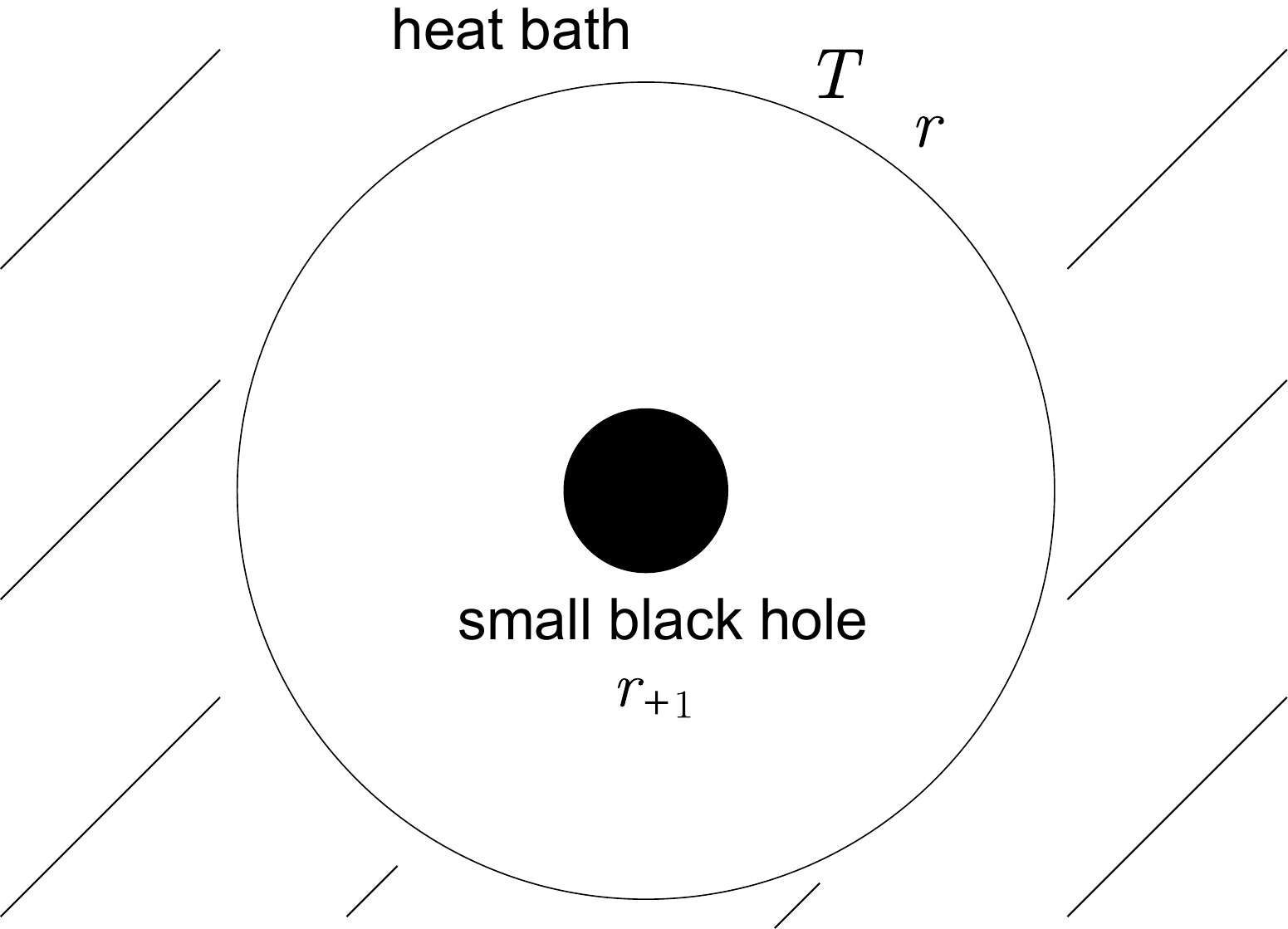}
\caption{
The small black hole solution $r_{+1}$ is depicted
inside the heat bath cavity characterized by its radius $r$
and temperature $T$.
}
\label{smallbhheatbath}
\end{figure}

For the smaller black hole $r_{+1}$, see Fig.~\ref{smallbhheatbath},
let us write the general form
of the solution as a Taylor expansion on $\pi r T$
around ${r_+}=0$.
Let us call $r_{+1}$ the gravitational
radius of the smaller black hole. Here, we write 
$r_{+1} =r_{+1}(\pi r T)$ as
\begin{equation}\label{expansion1}
r_{+1}=r
\left(
\sum_{i=1}^\infty
\frac{a_i}{(\pi r T)^i}
\right)\, ,
\end{equation}
where the $a_i$ are constants to be determined. 
Now, 
we need the expanding expressions for 
$\left(\frac{r_{+1}}r \right)^{d-1}$ and
$\left(\frac{r_{+1}}r \right)^2$
so that each power in $\pi r T$ cancels out in
Eq.~(\ref{polynomial1}).
Using Eq.~(\ref{expansion1})
we find 
$\left(\frac{r_{+1}}r \right)^{d-1}= 
\frac{a_1^{d-1}}{\left( \pi r T \right)^{d-1}} + 
\frac{(d-1) a_1^{d-2}a_2} {\left( \pi r T \right)^{d}} 
+ ...+\frac{2a_1a_{d-2}}{(\pi r T)^{d-1}}+...$ and
$\left(\frac{r_{+1}}r \right)^2=\frac{a_1^2}
{\left( \pi r T \right)^2} + 
\frac{2a_1a_2}{\left( \pi r T \right)^3} + ... $.
Due to the presence of the term 
$\left(\frac{d-3}{4\pi r T} \right)^2$ in Eq.~(\ref{polynomial1}),
$a_1$ can never be $0$, and since $d\geq4$, the lowest possible order
in
the $\left(\frac{r_{+1}}r \right)^{d-1}$ expansion
is in $\frac1{(\pi r T)^{3}}$, so, the only way
for the first term in
the 
$\left(\frac{r_{+1}}r \right)^2$ expansion
to cancel out in
Eq.~(\ref{polynomial1}) is by setting
$a_1 = \frac{d-3}{4}$.
Moreover, 
since the lowest order showing up
in
the $\left(\frac{r_{+1}}r \right)^{d-1}$ expansion
is in
$\frac1{(\pi r T)^{d-1}}$, then all
orders between $-1$ and $-(d-2)$ must
cancel out by setting the respective coefficient to zero, i.e.,
$a_i = 0$ for $2 \leq i \leq d-3$.
Thus, the next order term
in the expansion is that of $a_{d-2}$.
This is because by taking into account
that $a_i = 0$ for $2 \leq i \leq d-3$, 
the term in
$\frac1{(\pi r T)^{d-1}}$ arising from the squared term
$\left(\frac{r_+}{r} \right)^2$ in Eq.~(\ref{polynomial1})
is
$\frac{2a_1a_{d-2}}{(\pi r T)^{d-1}}$.
This term, along with the one coming
from the term in
the $\left(\frac{r_{+1}}r \right)^{d-1}$
expansion, will have to cancel out
in Eq.~(\ref{polynomial1})
by setting
$a_{d-2}=\frac{1}{2}\left( \frac{d-3}{4} \right)^{d-2}$.
With these values for the $a_i$, Eq.~(\ref{expansion1})
is now 
\begin{equation}\label{bh1app}
r_{+1} = r \left( \frac{d-3}{4 \pi r T} +
\frac12 \left(\frac{d-3}{4 \pi r T}\right)^{d-2} + 
\mathcal O\left(\frac1{(\pi r T)^{d-1}}\right) \right)
\end{equation}

\begin{figure}[h]
\centering
\includegraphics[width=0.20\textwidth]{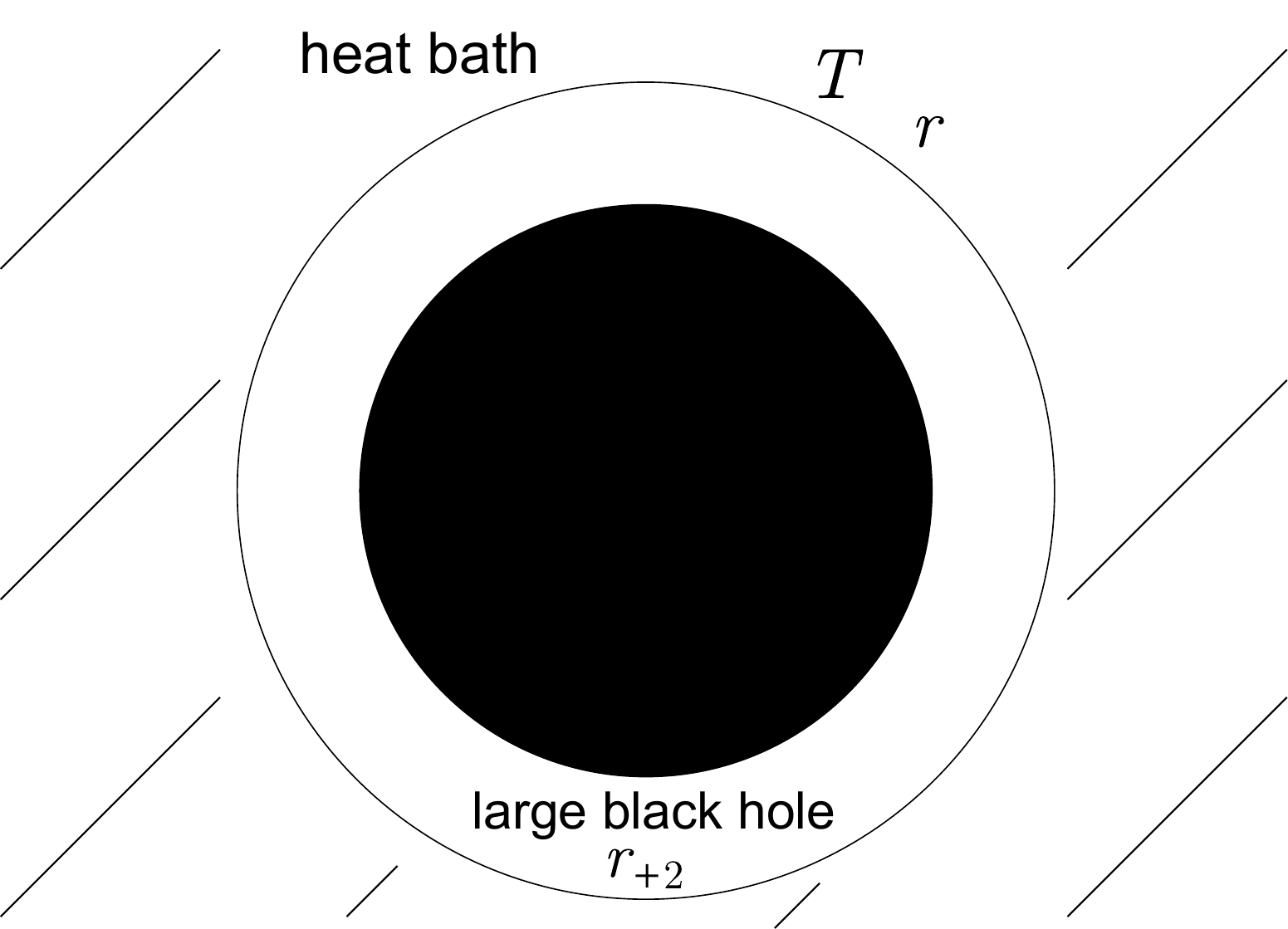}
\caption{
The large black hole solution $r_{+2}$ is depicted
inside the heat bath cavity characterized by its radius $r$
and temperature $T$.
}
\label{largebhheatbath}
\end{figure}

For the larger black hole $r_{+2}$, see Fig.~\ref{largebhheatbath},
let us write the general form
of the solution as a Taylor expansion on $\pi r T$
around $r$.
Let us call $r_{+2}$ the gravitational
radius of the larger black hole. Here, we write 
$r_{+2} =r_{+2}(\pi r T)$ as
\begin{equation}\label{expansion2}
r_{+2}=r
\left(
\sum_{i=0}^\infty
\frac{b_i}{(\pi r T)^i}
\right)\, ,
\end{equation}
where the $b_i$ for $i\geq0$
are constants to be determined.
Since for $r_{+2}$ the expansion is around
$r$ one has $b_0=1$. 
Now, 
we need the expanding expressions for 
$\left(\frac{r_{+2}}r \right)^{d-1}$ and
$\left(\frac{r_{+2}}r \right)^2$
so that each power in $\pi r T$ cancels out in
Eq.~(\ref{polynomial1}).
Using Eq.~(\ref{expansion2})
we find
$\left(\frac{r_{+2}}r \right)^{d-1} = 1 + 
\frac{(d-1)b_1}{\pi r T} +
\frac{(d-1)\left(b_2+\frac{(d-2)b_1^2}{2}\right)}{(\pi r T)^2}$
and
$
\left(\frac{r_{+2}}r \right)^2 = 1 + \frac{2b_1}{\pi r T} + 
\frac{(b_1^2+2b_2)}{(\pi r T)^2} 
+ ...$.
Since the terms in $\frac1{\pi r T}$ of
the $\left(\frac{r_{+1}}r \right)^{d-1}$ expansion and the
$\left(\frac{d-3}{4\pi r T} \right)^2$ expansion are only dependent on
$b_1$, and the polynomial Eq.~(\ref{polynomial1}) only has $\frac1{(\pi
r T)^2}$ showing up, the only way for these terms to cancel out is by
setting $b_1 = 0$.  Then, canceling out the terms in $\frac1{(\pi r
T)^2}$ is done by setting $b_2 = - \frac{d-3}{16}$.  One can now check
that the next leading order term is in $\frac1{(\pi r T)^4}$.  With
these values for the $b_i$, Eq.~(\ref{expansion2}) is now
\begin{equation}\label{bh2x}
r_{+2} = r\left(1 - \frac{d-3}{16\left( \pi r T\right)^2} 
+ \mathcal O\left( \frac1{(\pi r T)^4} \right) \right).
\end{equation}

Equations~(\ref{bh1app}) and~(\ref{bh2x}), are precisely the
Eqs.~(\ref{bh1}) and ~(\ref{bh2}), respectively, in the text.

\section{The photon orbit radius $r_{\rm ph}$
and the Buchdahl radius $r_{\rm Buch}$
in the $d$-dimensional Schwarzschild
solution}
\label{phbuch}

The Schwarzschild solution was generalized to $d$ dimensions by
Tangherlini \cite{tangherlini}, and it is variously called
$d$-dimensional Schwarzschild solution or Schwarzschild-Tangherlini
solution.  Here, we have opted to use the first name.

The photon orbit radius, or massless particle otbit radius, appears
naturally in the context of particle dynamics in a Schwarzschild
$d$-dimensional background.  In $d$ spacetime dimensions, it is
\cite{monteiro}
\begin{equation}\label{photonorbitradius}
r_{\rm ph}= \left(
\frac{d-1}{2} \right)^\frac1{d-3}r_+\,. 
\end{equation}
For $d=4$, one gets $r_{\rm ph}=\frac32r_+$, a result which is derived
in all books in general relativity.  This radius is also the radius
for a cavity of radius $r$, below which a black hole with horizon
radius $r_+$ in the canonical ensemble is thermodynamically stable. At
this radius, $r=r_{\rm ph}$, the heat capacity $C_A$ at constant
ensemble area $A$, with $A=\Omega_{d-2} r^{d-2}$ and $\Omega_{d-2}$ being the
solid angle in a spherical $d$-dimensional spacetime, is zero,
$C_A=0$, and for $r<r_{\rm ph}$ the heat capacity is positive, $C_A>0$.

The Buchdahl radius appears naturally in the context of star structure
and dynamics in general relativity.  It is a bound that states
that under some generic conditions for a spherical star of radius $r$,
the spacetime is free of singularities for $r_{\rm Buch}\leq r$.  For
$r_{\rm Buch}\geq r$, the star supposedly can collapse into a black
hole.  In $d$ spacetime dimensions, $r_{\rm Buch}\geq r$ is
\cite{wright}
\begin{equation}\label{buchdahlradius}
r_{\rm Buch}=\left(
\frac{ (d-1)^2 }{ 4(d-2) } \right)^{\frac{1}{d-3}}r_+\,,
\end{equation}
This radius is as well a limit for spherical thin shells in
$d$-dimensional spacetimes that have an equation of state given by
$p\leq\frac{1}{d-2}\sigma$, where $\sigma$ is the energy density of
the shell and $p$ the tangential pressure on the shell. At the
equality one gets the Buchdahl radius given in
Eq.~(\ref{buchdahlradius}).  In such a case, $\sigma = \frac{1}{4\pi}
\frac{d-2}{d-1} \frac{1}{r_{\rm Buch}}$ and $p = \frac{1}{4\pi}
\frac{1}{d-1} \frac{1}{r_{\rm Buch}}$.  Shells with lesser radius have
to have a stiffer equation of state.  So $p\leq \frac{1}{d-2}\sigma$
imposes a Buchdahl bound for shells.  For $d=4$ one finds $r_{\rm
Buch}=\frac98r_+$, a result derived by Buchdahl first for a spherical
perfect fluid star \cite{buchdahl} and later for thin shells in
\cite{andreasson}.  The Buchdahl radius is also the radius that a
cavity in the canonical ensemble for spherical gravitation has, such
that the free energy $F$ of the system is zero $F=0$, and so
above the Buchdahl radius, classical
hot flat space does not nucleate into a
black hole, and
below the Buchdahl radius classical hot flat space does nucleate
into a black hole. It is clear that the two contexts in which $r_{\rm
Buch}$ appears are clearly correlated.  That the Buchdahl radius
enters into thermodynamics of black holes in the canonical ensemble
was noticed first in \cite{andrelemosd5} where in five dimensions the
radius is $r_{\rm ph}=\frac2{\sqrt3}r_+$.

Thus, the photon orbit radius $r_{\rm ph}$ and the Buchdahl radius
$r_{\rm Buch}$ appear in two separate contexts, the former in both
particle dynamics and in thermodynamics and the latter in both star
dynamics and in thermodynamics.  The two contexts for $r_{\rm ph}$,
precisely,
particle dynamics in a Schwarzschild background on one side and black
hole thermodynamic stability on the other, are somehow correlated,
although this correlation has not been clearly interpreted.  The two
contexts for $r_{\rm Buch}$, namely,
general relativistic star solutions and
gravitational collapse dynamics on one side and black hole
thermodynamic on the other, are, on the other hand, clearly correlated.
It also hints that $r_{\rm Buch}$ is an intrinsic property of the
Schwarzschild spacetime, as the radius of the photon orbit, $r_{\rm
ph}$, is.  Note also from Eqs.~(\ref{photonorbitradius})
and~(\ref{buchdahlradius}) that $\frac{r_{\rm Buch}}{r_{\rm
ph}}=\left(\frac12\frac{d-1}{d-2}\right)^\frac{1}{d-3}$.  For
$d\to\infty$ one has $r_{\rm Buch}=r_{\rm ph}=r_+=0$.


\section{Connection to thermodynamics of thin shells
in $d$ spacetime dimensions}
\label{secshells}

Besides the black hole in the canonical ensemble, another system that
can have an exact thermodynamic treatment is provided by spherical
thin shells.  We  compare here
the thermodynamics of Schwarzschild black
holes
and classical hot flat
space in a cavity with radius $r$ at a fixed temperature $T$ in the
canonical ensemble in $d$ dimensions that we 
analyzed with the thermodynamics of a
self-gravitating Schwarzschild thin shell, i.e., a
thin shell with a Minkowski interior
and a Schwarzschild exterior, with radius
$r$ at a fixed temperature $T$
in $d$ dimensions~\cite{dshells}.  These thin matter shells are
$(d-2)$-dimensional branes in a spacetime of $d$ dimensions.

In the black hole in the canonical ensemble, case one has 
a cavity bounded by a massless boundary or massless thin shell, which
has radius $r$ and is at temperature $T$. The black
hole, when there is one,
is inside the boundary, and it has a gravitational or
event horizon radius $r_+$. There is also the possibility
that inside the cavity, there is only hot flat space,
which for this purpose is pure hot Minkowski space, i.e.,
classical hot flat space.
In the self-gravitating thin matter shell case, one has
that the shell is located at radius $r$
and is at fixed temperature $T$. The shell has rest mass $M$,
and so the spacetime has a gravitational
radius $r_+$, which is not an event horizon radius,
since there is no event horizon in this case. The thin shell
is a classical object.

Let us analyze the procedures for a black hole in the canonical
ensemble in $d$ dimensions and the procedure for the thermodynamic
thin matter shell in $d$ dimensions. The procedures are different. The
procedure for the black hole in a cavity is through the path integral
statistical mechanics approach  where a gravitational canonical
ensemble is defined which is then used to obtain all the
thermodynamic properties, as we have seen here for $d$-dimensional
spacetimes.

The procedure for the self-gravitating thin shell is
through local thermodynamics alone. The first law of thermodynamics at
the thin shell is used.  Let us see this, see also~\cite{dshells} for
a thorough analysis of thermodynamics of thin shells in $d$ spacetime
dimensions. In the thermodynamic analysis of a Schwarzschild thin
matter shell, a spherical static matter shell with rest mass $M$,
radius $r$, thus area $A=\Omega_{d-2} r^{d-2}$, and tangential pressure
$p$, with a well-defined local temperature $T$, obeys the first law of
thermodynamics $TdS = dM + pdA$, where $S$ is its entropy. $T$ and $p$
have to be provided through equations of state, and then the entropy
is generically given by $S=S(M,A)$. Using the spacetime general
relativity junction conditions one gets a relation between the
gravitational radius $r_+$, the proper mass $M$, and $r$, i.e.,
$r_+=r_+(M,r)$, and in addition an expression for the tangential
pressure $p$ in terms of $M$ and $A$.  Another set of conditions
besides the junction conditions is the one provided by the
integrability conditions for the first law, so that the entropy $S$ is
an exact differential.
For a Schwarzschild shell there is only one integrability condition.
It gives that the local temperature at the shell $T(M,r)$, or
$T\left(r_+(M,r),r\right)$ if one prefers, must have the Tolman form
for the temperature, i.e., $T(r_+,r) = \frac{T_\infty(r_+)}{k(r_+,r)}$
where $k(r_+,r)$ is the redshift factor, $k=\sqrt{1-\frac{r_+}{r}}$,
and $T_\infty(r_+)$ is a function of $r_+$ only to be chosen at our
will.  Indeed, $T_\infty(r_+)$ is a free function. Physically,
$T_\infty(r_+)$ can be interpreted as the temperature a small amount
of radiation would have at infinity after leaking out from the shell
at temperature $T$.

Let us analyze now the results for a black hole in the canonical
ensemble in $d$ dimensions and the results for the thermodynamic thin
matter shell in $d$ dimensions. The results have many similarities.

First, we analyze and compare the temperatures in each case.  For the
black hole in the canonical ensemble, the temperature $T$ of a heat
bath at the cavity's boundary at radius $r$ is fixed, and since the
black hole has mandatorily the Hawking temperature $T_H$, this
obliges, through the Tolman formula, the black hole radius to be
fixed, the computation showing that there are two equilibrium black
hole solutions, one large and stable and one small and unstable.  For
the thermodynamic thin matter shell at radius $r$, one puts it at some
fixed temperature $T$ which it is shown to obey the strict Tolman
formula, $T (r_+,r) = \frac{T_\infty(r_+)}{k(r,r_+)}$, with
$T_\infty(r_+)$ a free function.  This free function can be any
well-behaved function of $r_+$. In particular, $T_\infty(r_+)$ can
have the Hawking expression $T_\infty(r_+)=T_H$.  In this case, when
the thin shell has its temperature at infinity equal to the Hawking
temperature, then the two systems, namely, the black hole in the
canonical ensemble and the thin shell, are thermodynamically identical
in many respects.

Second, we analyze and compare the energies and pressures in each
case.  For the black hole in the canonical ensemble the thermodynamic
energy $E$ at the cavity's radius $r$ is $E =
\frac{(d-2)\Omega_{d-2}\,r^{d-3}}{8\pi} \left(1- \sqrt{1-
\frac{r_+^{d-3}}{r^{d-3}}}\right)$, and so with
$r_+^{d-3}=\frac{16\pi}{(d-2)\Omega_{d-2}}m$ one finds that the
spacetime mass $m$ is $m = E - \frac{4\pi E^2}{(d-2)\Omega_{d-2}\,
r^{d-3}}$.
For the black hole in the canonical ensemble the pressure $p$, which
is a thermodynamic tangential pressure, at the cavity's radius $r$, is
$p = \frac{d-3}{16\pi r \sqrt{1-\frac{r_+^{d-3}}{r^{d-3}}}} \left( 1 -
\sqrt{1-\frac{r_+^{d-3}}{r^{d-3}} }\right)^2$.
For the
thin matter shell at radius $r$,
assumed to be composed of a perfect fluid,
one has to find its stress-energy tensor $S_{ab}$,
where $a,b$ are spacetime
indices on the shell.
$S_{ab}$ can be put in diagonal form
and its components are characterized by
the rest mass energy density $\sigma$
and the  tangential
pressure $p$ acting on  a $(d - 2)$-sphere at radius $r$.
The junction conditions give that the
rest mass energy density $\sigma$
is  $\sigma=\frac{(d-2)}{8\pi r} \left(1- \sqrt{1-
\frac{r_+^{d-3}}{r^{d-3}}}\right)$.
Since the rest mass $M$ of the shell is given 
by
$M=\sigma\,A$ with $A=\Omega_{d-2} r^{d-2}$,
the rest mass  is $M =
\frac{(d-2)\Omega_{d-2}\,r^{d-3}}{8\pi} \left(1- \sqrt{1-
\frac{r_+^{d-3}}{r^{d-3}}}\right)$.
Putting 
$r_+^{d-3}=\frac{16\pi}{(d-2)\Omega_{d-2}}m$ into the expression for
$M$ one finds that the spacetime
mass $m$ is $m = M - \frac{4\pi M^2}{(d-2)\Omega_{d-2}\, r^{d-3}}$.
For the thin matter shell at radius $r$ the pressure $p$, which is a
dynamical tangential pressure derived from the junction conditions, is
$p = \frac{d-3}{16\pi r \sqrt{1-\frac{r_+^{d-3}}{r^{d-3}}}} \left( 1 -
\sqrt{1-\frac{r_+^{d-3}}{r^{d-3}} }\right)^2$.
Clearly, the thermodynamic energy $E$ in the black hole case
and the rest
mass $M$ in the thin shell case have the same expression and so can be
identified, i.e., $E=M$. $E$ and $M$ are
quasilocal energies.
Also, clearly, the
thermodynamic pressure $p$ in the black hole case and the dynamical
pressure $p$ in the thin shell have the same expression and so can be
identified.

Third, we analyze and compare
the entropies in each case.  For the black hole in
the canonical ensemble, the entropy is the Bekenstein-Hawking area law
$S=\frac14 A_+$, for both the stable and the unstable black holes.  For
the thermodynamic thin matter shell one finds that for any
well-behaved $T_\infty(r_+)$ its entropy is given by a function of
$r_+$ alone, $S=S(r_+)$,
independent of the shell radius $r$.
In particular, when the shell is put at
a temperature $T$ such that the temperature at infinity is the
Hawking temperature $T_\infty(r_+)=T_H$, then the entropy
of the shell $S=S(r_+)$ is definitely given by the Bekenstein-Hawking
area law $S=\frac14 A_+$. Moreover, when the shell is at $r_+$,
$r=r_+$, then the temperature at infinity
has to be mandatorily the Hawking
temperature, otherwise quantum effects render the whole system
unstable and undefined. Thus, when the shell turns into a black hole,
more properly into a quasiblack hole, one recovers from the shell
thermodynamics the black hole's expressions.

Fourth, we analyze and compare
the thermal stability in each case.  For the black
hole in the canonical ensemble the heat capacity $C_A$ is the quantity
that signals thermodynamic stability if $C_A \geq 0$ from
thermodynamic instability if $C_A < 0$. It was shown that $C_A =
\frac{(d-2)}{2(d-1)} \Omega_{d-2} r_+r^{d-3} \frac{
1-\frac{r_+^{d-3}}{r^{d-3}} } { 1 - \frac{2}{d-1}
\frac{r^{d-3}}{r_+^{d-3}} }$, and it implies that when the cavity's
radius $r$ is less than or equal to the radius of the circular photon
orbits, i.e., $r_{+}<r \leq r_{\rm ph}$ the black hole is
thermodynamically stable, otherwise unstable, this meaning that the
large black hole $r_{+2}$ is the stable one and the smaller $r_{+1}$ is
unstable.
For the thin matter shell there is also the thermodynamic stability
criterion $C_A \geq 0$, as well as other stability criteria which
further restrict the thermodynamic stability. The particular
interesting case, the one related to the black hole in the canonical
ensemble, is when the temperature
of the shell at infinity is
the Hawking temperature $T_H$. In this
very case the heat capacity $C_A$ has the expression $C_A =
\frac{(d-2)}{2(d-1)} \Omega_{d-2} r_+r^{d-3} \frac{
1-\frac{r_+^{d-3}}{r^{d-3}} } { 1 - \frac{2}{d-1}
\frac{r^{d-3}}{r_+^{d-3}} }$ and so for stability the self-gravitating
matter shell must be placed between its own gravitational and its
photon sphere for stability, i.e., $r_{+}<r \leq r_{\rm ph}$.
Thus, in the case that the temperature of
the shell at infinity is the Hawking temperature, 
and so in the situation that is thermodynamic
similar to the black hole,
the thermodynamic criterion of positive heat capacity
gives the same result for both systems.

Fifth, we analyze and compare
the generalized free energy function in each case.
For the black hole in the canonical ensemble, the free energy $F$ gives
a special cavity radius $r$ for which it is zero. This radius is the
Buchdahl radius $r_{\rm Buch}$ that appears naturally in
general relativistic star
structure and dynamics, especially in star gravitational collapse.
It also appears in
the black hole thermodynamic context.  For $r\geq r_{\rm Buch}$ classical hot
flat space that does transition to a black hole, for $r< r_{\rm Buch}$,
there is a phase transition from classical hot flat space to a black
hole.
For the thin matter shell, with the identification of the mass $M$
with the thermal energy $E$, $M=E$, a free energy $F$ can be defined
by $F=M-TS$.  When the temperature of the shell at infinity is the
Hawking temperature $T_H$ and thus the shell has the Bekenstein-Hawking
entropy, such a free energy $F$ also gives the Buchdahl radius $r_{\rm
Buch}$ as the special cavity radius $r$ for which $F=0$, presumably
meaning that it is
energetically favorable for the shell to disperse away
at this radius in the given conditions. Here, the Buchdahl radius
$r_{\rm Buch}$ appears also as a structure and dynamic radius 
on top of being a
thermodynamic one.  Indeed, by imposing that the equation of state for
the matter in the shell obeys $p\leq\frac{1}{d-2}\sigma$, i.e., by
imposing that the pressure is equal to or less than the
radiation pressure, a
sort of energy condition, specifically, the trace of
the stress-energy tensor $S_{ab}$ is equal to or less than
zero, ${\rm Tr}\, S_{ab}\leq0$, one finds that the bound
$p=\frac{1}{d-2}\sigma$, is satisfied for the $d$-dimensional
Buchdahl radius, i.e., $r_{\rm Buch}= \left( \frac{ (d-1)^2 }{
4(d-2) } \right)^{\frac{1}{d-3}}\,{r_+}$.  Shells with lesser radius
have to have a stiffer equation of state.  So, $p\leq
\frac{1}{d-2}\sigma$ imposes a Buchdahl bound for shells.  Thus, one
finds that the free energy in both cases is zero when the radius of
the cavity $r$ or the radius of the shell $r$ are at the Buchdahl
radius, in the latter case meaning that the pressure at the shell is
equal to the radiation pressure.

Thus, this thorough comparison between the black hole in the canonical
ensemble in $d$ dimensions and the thin matter shell in $d$ dimensions
shows that indeed, when the situations are similar, explicitly, when
the shell's temperature at infinity is the Hawking temperature, and
for the quantities that it makes sense to perform a comparison, the two
systems behave thermodynamically in similar ways. The
boundary of the black hole
cavity at a definite temperature defines a heat reservoir,
analogously, the shell at a definite temperature is a heat reservoir.

\section{Quantum hot flat space in $d$ spacetime dimensions}
\label{secApB}

The first law of thermodynamics
for quantum hot flat space is written as 
\begin{equation}\label{1stlawhfs}
TdS_{\rm HFS} = dE_{\rm HFS} + P_{\rm HFS} dV\,,
\end{equation}
where $T$ is the temperature of the space, 
$S_{\rm HFS}$ is the quantum hot flat space
entropy, 
$E_{\rm HFS}$ is its internal energy,
$P_{\rm HFS}$ is its radiation
pressure, and $V$ is the volume
it occupies.
The internal energy $E_{\rm HFS}$ of
such a radiation gas has the usual formula, definitely,
\begin{equation}\label{EnergyHFS}
E_{\rm HFS} = N V a T^d,
\end{equation}
where $N$ is the total number of
massless states,
and 
$a$ is a
quantum mechanics constant given by
$
a = \frac{\Gamma(d)\zeta(d)}{2^{d-2}\pi^{\frac{d-1}{2}}\Gamma 
\left( \frac {d-1}2 \right)}
$,
with $\Gamma$ and $\zeta$ being the gamma and zeta functions,
respectively.  The constant $a$ is related to the $d$-dimensional
Stefan-Boltzmann constant $\sigma$ through $\sigma=a\, \frac{\sqrt \pi
\,\, \Gamma \left( \frac d 2 \right)} {2 \, \Gamma \left( \frac
{d-1}{2} \right)}$~\cite{landsberg}, wich
for $d=4$ simplifies to
$\sigma=a$.
The equation of state for radiation that gives
a relation between the radiation pressure  $P_{\rm
HFS}$, $V$, and $E_{\rm HFS}$, is
\begin{equation}\label{RadiationHFSpressure}
P_{\rm HFS}\,V=
\frac{1}{d-1}
E_{\rm HFS} \,,
\end{equation}  
so that using
Eq.~(\ref{EnergyHFS}), one finds
$P_{\rm HFS}= \frac{N  a}{d-1} T^d$.
From the first law Eq.~(\ref{1stlawhfs})
and using Eqs.~(\ref{EnergyHFS})
and~(\ref{RadiationHFSpressure}), one finds
the entropy of quantum hot flat space
\begin{equation}\label{RadiationHFS}
S_{\rm HFS} = \frac{d}{d-1}
\frac{ E_{\rm HFS}}{T} \,,
\end{equation}  
i.e., $S_{\rm HFS}=\frac{d}{d-1}N V a T^{d-1}$.
The free energy for quantum hot flat space is
\begin{equation}\label{freeenergyHFS}
F_{\rm HFS}=E_{\rm HFS}-TS_{\rm HFS}\,.
\end{equation}
From Eq.~(\ref{RadiationHFS}) on
 Eq.~(\ref{freeenergyHFS})
one has
\begin{equation}\label{freeenergyHFS2}
F_{\rm HFS}=-\frac{1}{d-1}E_{\rm HFS}\,,
\end{equation}
and using Eq.~(\ref{EnergyHFS}),
one finds
$F_{\rm HFS}=-\frac{1}{d-1}N V a T^d$.
Now, the volume $V$ of a spherical
cavity in $d$ dimensions is
$V=\frac{\Omega_{d-2}}{d-1} r^{d-1}$,
with $\Omega_{d-2} =
\frac{2\pi^{\frac{d-1}{2}}}{\Gamma\left(\frac{d-1}{2}\right)}$
being the solid
angle in spherical $d$-dimensional spacetime.
We have not specified yet
the number of degrees of
freedom $N$, and as it is, the expression works for any
radiation gas of massless particles in
flat space at finite temperature in $d$ dimensions.
If there are only gravitons within the cavity, $N$ is
given by $N = \frac{d(d-3)}{2}$.
For this $N$, Eq.~(\ref{freeenergyHFS2})
together with Eq.~(\ref{EnergyHFS}) gives
\begin{equation}\label{FHFSfinal}
F_{\rm HFS} = - \frac{d(d-3)}{2(d-1)^2}
a\,\Omega_{d-2} r^{d-1} T^d\,,
\end{equation}
which is the expression we use in Eq.~(\ref{achfs}).
To complete,
since $I=\beta F$ and $\beta=\frac1T$, the  action for 
$d$-dimensional quantum hot flat space is
\begin{equation}\label{APachfs}
I_{\rm HFS} =  - \frac{d(d-3)}{2(d-1)^2}
a\,\Omega_{d-2} r^{d-1} T^{d-1}.
\end{equation}

\section{Classical hot flat space in $d$ spacetime dimensions as 
a product of quantum hot flat space and the corresponding black hole
phase transitions}
\label{appchfs}
\setcounter{figure}{0}

It is of interest
to understand the passage
to classical hot flat space in $d$ spacetime dimensions
from quantum hot flat space and look
into the 
black hole phase transitions
from classical hot flat space in some more detail.

\begin{figure}[h]
\centering
\includegraphics[width=0.44\textwidth]{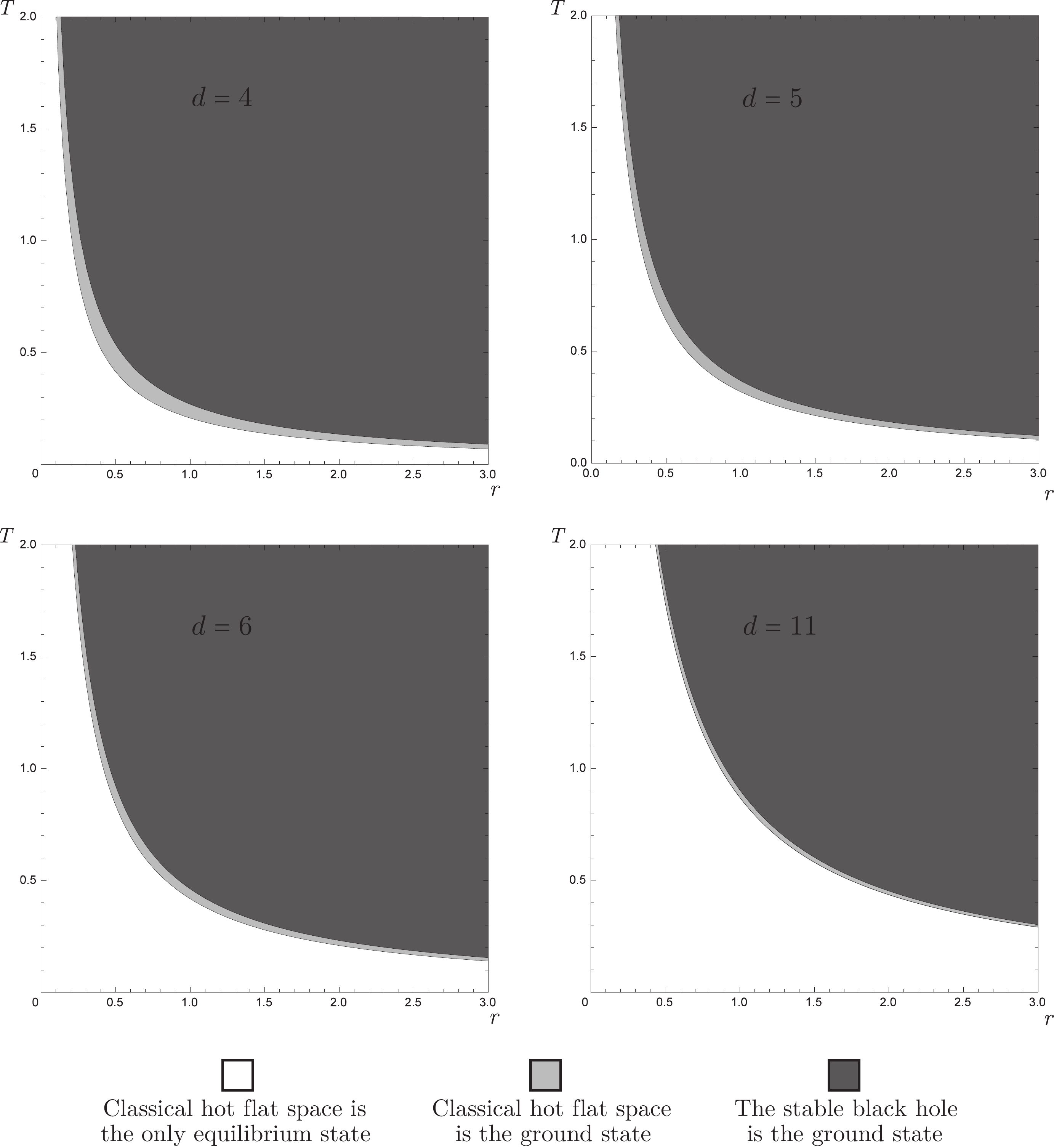}
\caption{
Phase diagram of the gravitational canonical ensemble in the classical
hot flat space case with a plot of the cavity radius $r$, in arbitrary
units, versus the temperature $T$, in arbitrary units, for four
different dimensions, $d=4$, $5$, $6$, and $11$.  Each point in the
plots represents a different configuration of the ensemble.  Each plot
is separated in three different regions.  The white region in each
plot, delimited by the hyperbola $\pi r T \geq \frac{d-3}{4}\left[
\left( \frac{2}{d-1} \right)^{\frac{2}{d-3}} -\left( \frac{2}{d-1}
\right)^{\frac{d-1}{d-3}}\right]^{-1/2}$, see Eq.~(\ref{bhcondition}),
is the region where in thermodynamic
equilibrium there is 
only classical hot flat space, 
with no black holes.  Each point in the
hyperbola gives the photon radius $r_{\rm ph}$ and the corresponding
temperature.  The gray region in each plot is characterized by having
as the ground state classical hot flat space and is delimited also by
the other hyperbola $\pi r T= \left(\frac{(d-1)^{d-1}}{4^{d-2}(d-2)}
\right)^{\frac{1}{d-3}}$, see Eq.~(\ref{fullc}). Each point in this
hyperbola gives the Buchdahl radius $r_{\rm Buch}$ and the
corresponding temperature. The dark region in each plot, delimited by
the latter hyperbola, is characterized by having as the ground state
the larger stable black hole $r_{+2}$.
The thick black line in each plot yields a mixed phase,
i.e., a superposition of the
classical hot flat space ground state
phase with the
stable black holes ground state
phase.
Comparing this figure with
Fig.~\ref{GroundStates} for quantum hot flat space, one sees that
classical hot flat space approximates quantum hot flat space for large
cavity radius $r$ and low temperature $T$.
}
\label{GroundStatesclassical}
\end{figure}

In classical hot flat space, one puts $a=0$, and so Eq.~(\ref{achfs}),
or Eq.~(\ref{FHFSfinal}), reads now
\begin{equation}\label{chfs5}
F_{\rm HFS} = 0\,.
\end{equation}
Thus, Eq.~(\ref{fr+2versushfs}), which states the condition for the
stable black hole $r_{+2}$ to be the ground state, turns into
\begin{equation}\label{classhotbh}
F(r_{+2})\leq 0\,.
\end{equation}
The phase diagram for black holes and a classical hot flat space in $d
= 4$, 5, 6, and 11 dimensions, is now given in
Fig.~\ref{GroundStatesclassical}, which is the limit of
Fig.~\ref{GroundStates} when $a=0$.

Equation~(\ref{classhotbh}) has no extrema, unlike the quantum case of
Eq.~(\ref{fr+2versushfs}).  Using Eq.~(\ref{fr+2versushfs}), we have
found a minimum radius $r_\text{min}$ such that if the cavity's radius $r$
obeys $r<r_\text{min}$, only quantum hot flat space can be the ground state.  We
have also found, using Eq.~(\ref{fr+2versushfs}), a maximum
temperature $T_\text{max}$ such that if the cavity's temperature $T$ obeys
$T>T_\text{max}$, only quantum hot flat space could be the
ground state.  For
classical hot flat space, these extrema do not occur, or to be more
precise, one finds $r_\text{min}=0$ and $T_\text{max}=\infty$.

Since $r_\text{min}$ and $T_\text{max}$ do
not enter the problem if one uses 
classical
hot flat space, the only parameter that matters is $\pi r T$, see
Eq.~(\ref{fullc}) or Eq.~(\ref{F<0}).  Thus, here there are two
important hyperbolas when working out the phase diagram for classical
hot flat space. There is the hyperbola $\pi r T = \frac{d-3}{4}\left[
\left( \frac{2}{d-1} \right)^{\frac{2}{d-3}} -\left( \frac{2}{d-1}
\right)^{\frac{d-1}{d-3}}\right]^{-1/2}$, see Eq.~(\ref{bhcondition}),
that separates the region where classical hot flat space is the only
equilibrium state from the region where classical hot flat space is
the ground state.
As we have seen, each point in this hyperbola gives
the photon radius $r_{\rm ph}= \left( \frac{d-1}{2}
\right)^\frac1{d-3}r_+$ for the radius of the cavity and the
corresponding cavity's temperature.  There is the other hyperbola $\pi
r T= \left(\frac{(d-1)^{d-1}}{4^{d-2}(d-2)} \right)^{\frac{1}{d-3}}$,
see Eq.~(\ref{fullc}), that separates the region where classical hot
flat space is the ground state from the region where stable black
holes $r_{+2}$ is the ground state and so classical hot flat space can
nucleate stable black holes $r_{+2}$. 
Each point in this hyperbola, characterized by a specific 
cavity radius and a specific cavity's temperature, 
corresponds to a specific thermodynamic system, and, as 
we have seen, each such point also yields the Buchdahl 
radius,
$r_{\rm Buch}=\left( \frac{
(d-1)^2 }{ 4(d-2) } \right)^{\frac{1}{d-3}}r_+$,
for the system under consideration.

Thus, there are three phases.  One phase when the cavity's radius $r$
and the cavity's temperature $T$ only give the possibility of the
existence of classical hot flat space, there are no stable equilibrium
black holes $r_{+2}$, and for that matter also no unstable equilibrium
black holes $r_{+1}$, but eventually black holes out of thermodynamic
equilibrium may appear in this phase.  Another phase is when the
cavity's radius $r$ obeys $r_\text{Buch}< r < r_\text{ph}$, where
classical hot flat space is the ground state and so stable black holes
$r_{+2}$ can transition into classical hot flat space.  And yet
another phase is when the cavity's radius $r$ obeys $r <
r_\text{Buch}$ where the stable black hole $r_{+2}$ is the ground
state and so classical hot flat space can nucleate stable black holes
$r_{+2}$.  There is also a mixed phase, which is a superposition of
the two previous phases.  In Fig.~\ref{GroundStatesclassical}, the
three phases are represented by the white, gray, and dark regions,
respectively, and the mixed phase is represented by a thick black line
between the gray and dark regions.

Comparing Fig.~\ref{GroundStatesclassical} for classical hot flat
space with Fig.~\ref{GroundStates} for quantum hot flat space, one sees
that classical hot flat space approximates quantum hot flat space for
large cavity radius $r$ and low temperature $T$.  Two important
consequences can be drawn from this comparison.  One consequence is
that in classical hot flat space, as the number of spacetime
dimensions increases, the region for which stable black holes can
transition into classical hot flat space gets smaller, whereas the
region for which classical hot flat space can nucleate stable black
holes gets larger, contrarily to what happens in quantum hot flat
space, thus showing clearly that the classical approximation is not
valid for a vast region of the $r\times T$ plane.  The other
consequence of this comparison is that the Buchdahl radius is an
important radius in the classical approximation, as one would expect.

\section{Synopsis and additions}
\label{conc}

The canonical ensemble statistical mechanics formalism has been used
for $d$-dimensional black holes, and the corresponding thermodynamics
has been studied in detail, thereby extending York's four-dimensional
results and previous five-dimensional results.

When working out the possible black hole radii for a given cavity
radius $r$ and a given cavity temperature $T$, one finds that in four
dimensions one needs to solve a cubic equation and in five dimensions
one needs to solve a quartic that can be reduced to a quadratic
equation.  In both cases, the solutions yield two real radii yielding
the two black holes.  In $d$ dimensions, one has a polynomial equation
of order $d-1$, and one could expect a different number of real
solutions, but there are also only two, with radii $r_{+1}$ and
$r_{+2}$, yielding again two possible black holes.

The formalism shows that for any dimension $d$, the two black holes in
equilibrium, the small $r_{+1}$ and the large $r_{+2}$, both have a
Bekenstein-Hawking entropy, although the procedure is only well
defined for the larger one since it is stable and the laws of
thermodynamics can be applied, whereas the smaller one is unstable and
cannot be treated properly through thermodynamics.

There are two distinct characteristic radii that appear automatically
naturally in the canonical black hole thermodynamics. One
characteristic radius is the photon sphere radius $r_{\rm ph}$. The
photon sphere radius that appears in the Schwarzschild geometry,
particularly when there is a black hole, is the radius for which any
massless particle, including photons, can have circular orbits.
Notably, in $d$ dimensions, for a given temperature $T$ of the heat
bath, the photon sphere radius also appears as separating black hole
systems that are thermodynamically stable from black hole systems that
are unstable.  Being a characteristic that seems to appear in any
dimension $d$, this shows that there is some intrinsic property of the
photon sphere location that connects it to thermodynamic
stability. However, a full explanation has not been given yet.  The
other characteristic radius that appears automatically in these
canonical ensemble thermodynamic systems is the Buchdahl radius
$r_{\rm Buch}$. The Buchdahl radius is the minimum radius that, under
certain general conditions, a spherically symmetric interior matter
solution with Schwarzschild exterior can have.  Surprisingly, this
radius also shows up in the canonical ensemble setting, as the radius
for which the free energy of the stable black hole $r_{+2}$ passes
through zero, as we have shown. This happens in any dimension
$d$. Since the two occasions at which $r_{\rm Buch}$ emerges are
totally distinct, it seems that $r_{\rm Buch}$ also signals some
intrinsic property of the spacetime geometry.  Indeed, $r_{\rm Buch}$
is also related to thin shells, where it appears both for dynamic and
for thermodynamic reasons.

The formalism also permits comparing the free energy of
$d$-dimensional quantum hot flat space with the free energy of the
$d$-dimensional stable black hole $r_{+2}$ and thereby identify the
conditions for which the ground state of the canonical ensemble is
quantum hot flat space or the stable black hole. It was found that for
sufficiently low cavity temperatures $T$ and sufficiently large cavity
radius $r$, in Planck units, the ground state is the stable black hole
phase, but in the limit of a very high number of dimensions $d$ this
phase gets vanishing small, and so in this limit black holes never
nucleate.


\end{document}